\documentclass[11pt,a4paper]{article}
\pdfoutput=1
\usepackage{jheppub}
\usepackage{indentfirst}
\usepackage{amsmath}
\usepackage{bigstrut}
\usepackage{amssymb}
\usepackage{array}
\usepackage{multirow}
\usepackage{floatrow}
\usepackage{etoolbox}
\usepackage{graphicx}
\usepackage{hyperref}
\usepackage{amsthm}
\usepackage{slashed}
\usepackage{array}
\usepackage{epstopdf}
\usepackage{graphicx}
\usepackage{fancyhdr}
\usepackage{listings}
\usepackage[table,xcdraw]{xcolor}
\usepackage{float}
\usepackage{subcaption}
\usepackage[table,xcdraw]{xcolor}
\usepackage{hyperref}
\usepackage{tikz}
\usetikzlibrary{shapes, arrows}

\usepackage[utf8]{inputenc}
\usepackage[T1]{fontenc}
\usepackage{lmodern}
\usepackage{textcomp}
\usepackage{microtype}
\usepackage[normalem]{ulem}

\usepackage{mathtools}
\usepackage{physics}
\usepackage{units}
\usepackage{siunitx}

\usepackage{graphicx}
\graphicspath{{figs/}}

\newcommand{\fig}[1]{Fig.~\ref{fig:#1}}

\usepackage{xcolor}

\preprint{FERMILAB-PUB-22-035-T}

\title{Detecting New Physics as Novelty\\
- Complementarity Matters }

\author[a]{Xu-Hui Jiang,}
\affiliation[a]{Department of Physics and Jockey Club Institute for Advanced Study, The Hong Kong University of Science and Technology, Hong Kong S.A.R., P.R.China}
\emailAdd{xjiangaj@connect.ust.hk}
\author[b,c]{Aurelio Juste,}
\affiliation[b]{Institut de F\'{i}sica d'Altes Energies (IFAE), Edifici Cn, Facultat de Ci\`{e}ncies, Universitat Aut\`{o}noma de Barcelona, E-08193 Bellaterra, Barcelona, Spain}
\affiliation[c]{Instituci\'{o} Catalana de Recerca i Estudis Avan\c{c}ats (ICREA), E-08010 Barcelona, Spain}
\emailAdd{Aurelio.Juste.Rozas@cern.ch}
\author[d]{Ying-Ying Li,}
\affiliation[d]{Theoretical Physics Department, Fermi National Accelerator Laboratory, PO Box 500, Batavia, IL 60510, U.S.A.}
\emailAdd{yingying@fnal.gov}
\author[a]{Tao Liu,}
\emailAdd{taoliu@ust.hk}

\abstract{Novelty detection is a task of machine learning that aims at detecting novel events without a prior knowledge. In particular, its techniques can be applied to detect unexpected signals from new phenomena at colliders. In this paper, we develop an analysis scheme that exploits the complementarity, originally studied in Ref.~\cite{Hajer:2018kqm}, between isolation-based and clustering-based novelty evaluators. This approach can significantly improve the performance and overall applicability of novelty detection at colliders, which we demonstrate using a variety of two dimensional Gaussian samples mimicking collider events. As a further proof of principle, we subsequently apply this scheme to the detection of two significantly different signals at the LHC featuring a $t\bar{t}\gamma\gamma$ final state: $t\bar t h$, giving a narrow resonance in the diphoton mass spectrum, and gravity-mediated supersymmetry, resulting in broad distributions at high transverse momentum. Compared to existing dedicated searches at the LHC, the sensitivities for detecting both signals are found to be encouraging.}

\begin{document}
\maketitle
\unitlength = 1mm

\section{Introduction}

Particle physics has a long history of applying Machine Learning (ML) techniques in data analyses~\cite{Guest:2018yhq}, where the complexity of event topologies at colliders post great challenges to traditional wisdom. For example, the neural network was applied to the top quark search by the D\O\ Collaboration in 90s~\cite{1996,PRLD0}, far before the deep neural network (DNN) became popularized due to hardware development and big-data availability about fifteen years ago~\cite{cios2017deep}. Another example is the Boosted Decision Tree (BDT)~\cite{boosting}. Although it was first introduced by the MiniBooNE Collaboration for the analysis of neutrino data in 2004~\cite{Roe:2004na}, BDTs are nowadays extensively used in the analysis of collider data. For both methods, the algorithms are first trained on labeled data, $i.e.$ in a supervised way, and then applied to classify testing data into the categories defined during training. The usage of these ML techniques in the analysis of collider events often result in significant improvements, assisting, e.g., in the discovery of the Higgs boson at Large Hadron Collider (LHC) in 2012~\cite{Aad:2012tfa, Chatrchyan:2012xdj}, and many other new phenomena searches and Standard Model (SM) measurements. 

The triumph of the Higgs-boson discovery propelled the ambition for new physics (NP) hunting at the energy frontier. Since then, many dedicated searches targeting well-motivated beyond-the-SM extensions such as supersymmetry (SUSY), Composite Higgs models, extra spatial dimensions, etc., are being pursued. Yet, no convincing signals of NP have been observed so far. This suggests that, if NP exists, it may present itself in a highly unexpected form. This strongly motivated the design of new analysis strategies that would allow NP to be detected in a more model-independent way and with a broader coverage in theory space, hence complementing the ongoing model-dependent search program at the LHC. 

In the science of ML, this involves a well-known task - novelty (or anomaly) detection, $i.e.$, detecting novel events without a prior knowledge. This implies that there is no data of the signal pattern available for model training. Yet, different from its usual applications such as the face- or fingerprint-recognition of a stranger, where individual novel event is expected to be detected, at colliders the novelty detection will be defined on a statistical basis. To address the task of novelty detection, a series of pioneering studies have been pursued in the last decades. Essentially, designing novelty evaluators formulated its mainline (for a review, see, e.g., Ref.~\cite{Pimentel14}). Depending on the characters of the events employed to perform this task, these novelty evaluators/algorithms can be roughly classified into two types~\cite{Hajer:2018kqm}~\footnote{In Ref.~\cite{Hajer:2018kqm}, we have demonstrated the isolation-based (${\mathcal O_{\rm iso}}$) and clustering-based (${\mathcal O_{\rm clu}}$) evaluators, together with the synergy-based evaluator (${\mathcal O_{\rm syn}}$) introduced below, using the $k$-nearest-neighbors ($k$NN)-based designs (${\mathcal O_{\rm trad}}$, ${\mathcal O_{\rm new}}$ and ${\mathcal O_{\rm comb}}$). In this paper, based on their nature, we will rename these designs as $k${\rm NN-based} ${\mathcal O_{\rm iso}}$, ${\mathcal O_{\rm clu}}$ and $ {\mathcal O_{\rm syn}}$, respectively.}~\footnote{The classification taken in this article is based on the event characters employed for evaluating the novelty of the collider events. The goal is to make use of the complementarity between these characters to further improve the quality of evaluation. Notably, multiple ways exist to classify the evaluators/algorithms of novelty detection. For example, one can also classify these evaluators/algorithms according to local binning versus global statistical test~\cite{DAgnolo:2018cun,DeSimone:2018efk}.}:
\begin{itemize}

\item Isolation-based (${\mathcal O_{\rm iso}}$). The novelty response for a given testing event is evaluated according to its isolation from the known-pattern data in the feature space. All of the other testing events are not directly involved in this evaluation. 

\item Clustering-based (${\mathcal O_{\rm clu}}$). The novelty response for a given testing event is evaluated according to the clustering around this event on top of the known-pattern data in the feature space. The other testing events, especially those nearby in the feature space, are potentially relevant in this evaluation.

\end{itemize}
Here the distribution of the known-pattern data or background events in both cases can be figured out by taking either Monte-Carlo simulation (semi-supervised ML) or data extrapolation (fully unsupervised ML).

\begin{table} 
 \begin{center}
  \begin{tabular}{|c|c|}
  \hline
  \multirow{3}{*}{${\mathcal O_{\rm iso}}$} & $k$-nearest-neighbors($k$NN)-based ${\mathcal O_{\rm iso}}$~\cite{Hajer:2018kqm} \\  
    \cline{2-2} & 
   Autoencoder(AE)-based~\cite{Heimel:2018mkt,Farina:2018fyg,Roy:2019jae,Cerri:2018anq,Cheng:2020dal,Knapp:2020dde,Blance:2019ibf,Bortolato:2021zic, Atkinson:2021nlt, Ngairangbam:2021yma} \\ 
    \cline{2-2} & Graph~\cite{Mullin:2019mmh}, classical $k$-means clustering~\cite{Buss:2022lxw} \\ \hline
      \multirow{6}{*}{${\mathcal O_{\rm clu}}$} 
&$k$NN-based ${\mathcal O_{\rm clu}}$~\cite{Hajer:2018kqm}, TS~\cite{DeSimone:2018efk}\\  \cline{2-2} & $t$-score~\cite{DAgnolo:2018cun, DAgnolo:2019vbw, Chakravarti:2021svb, dAgnolo:2021aun}, SOFIE~\cite{Aguilar-Saavedra:2021utu}, ANODE~\cite{Nachman:2020lpy}, Poissonian Mixture Model~\cite{Alvarez:2021zje} \\  \cline{2-2} & CWoLa~\cite{Metodiev:2017vrx, Collins:2018epr, Collins:2019jip, Aad:2020cws}, TNT~\cite{Amram:2020ykb}, SALAD~\cite{Andreassen:2020nkr}  \\  \cline{2-2} & SULU~\cite{Dahbi:2020zjw} \\ \cline{2-2} & UCluster~\cite{Mikuni:2020qds} 
\\ \hline
\end{tabular}
\end{center}
\caption{A short summary of the isolation-based and clustering-based novelty evaluators/algorithms.}
\label{tab:algorithms}
\end{table}

A short summary of the evaluators/algorithms proposed in the literature so far for novelty detection at colliders is given in Table~\ref{tab:algorithms}. In this landscape, the autoencoder (AE) was first introduced in Ref.~\cite{Hajer:2018kqm}, where two $k$NN-based measures, ${\mathcal O_{\rm iso}}$ and ${\mathcal O_{\rm clu}}$, were proposed for novelty evaluation in the AE latent space. After that, a set of metrics based on an ordinary AE~\cite{Heimel:2018mkt, Farina:2018fyg, Roy:2019jae}, a variational AE~\cite{Cerri:2018anq, Cheng:2020dal, Bortolato:2021zic}, an adversarial AE~\cite{Heimel:2018mkt, Blance:2019ibf, Knapp:2020dde}, and a graph AE~\cite{Atkinson:2021nlt} were suggested as novelty measures. They mainly include reconstruction error in Ref.~\cite{Heimel:2018mkt, Farina:2018fyg, Roy:2019jae, Cerri:2018anq, Cheng:2020dal, Knapp:2020dde, Blance:2019ibf, Atkinson:2021nlt}, Kullback-Leibler divergence~\cite{KL-D} in Ref.~\cite{Cheng:2020dal, Bortolato:2021zic}, Energy Mover's Distance~\cite{Komiske:2019fks} in Ref.~\cite{Cheng:2020dal}, and Logits and feature scores in Ref.~\cite{Knapp:2020dde}. Furthermore, the authors in~\cite{Ngairangbam:2021yma} studied the potential of novelty detection with variational-quantum-circuits-based quantum autoencoder. These proposals are clearly isolation-based, since the novelty response of each testing event is evaluated by them independently. In addition, a graph network was proposed in Ref.~\cite{Mullin:2019mmh} to search for SUSY events, whereby multiple distance metrics were used to obtain different distributions for the events following different topologies. Recently the authors of Ref.~\cite{Buss:2022lxw} studied anomalous jets at LHC which are far from the background clusters, using the distance of each testing event to the cluster centroids as a measure. 

The local density of each event was estimated with $k$NN in Refs.~\cite{Hajer:2018kqm, DeSimone:2018efk}, where the ${\mathcal O_{\rm clu}}$ evaluator in Ref.~\cite{Hajer:2018kqm} mimics the structure of $\frac{N-B}{\sqrt{B}}$ (where $N$ and $B$ denote the number of data and background events, respectively; see Sec.~\ref{sec:S2}) while a test statistic TS$(p_\text{test}, p_B)$ in Ref.~\cite{DeSimone:2018efk} was defined in the logarithmic form of the Neyman-Pearson lemma. The lemma was applied as a novelty score in Ref.~\cite{DAgnolo:2018cun, DAgnolo:2019vbw, Nachman:2020lpy} as well. In Ref.~\cite{DAgnolo:2018cun, DAgnolo:2019vbw, Chakravarti:2021svb, dAgnolo:2021aun}, the local density was parameterized straightforwardly with a supervised learning to distinguish data from the reference, with the loss function serving as a novelty score $t$. In Ref.~\cite{Aguilar-Saavedra:2021utu}, each event is evaluated based on the fraction of its $k$NN which are signal-like or have a pre-assigned ``signal'' label. ANODE~\cite{Nachman:2020lpy} utilizes the Neyman-Pearson lemma directly, but the physical distribution was simplified into a prior distribution via a masked autoregressive flow~\cite{NIPS2017_6c1da886}. The unsupervised tagging in~\cite{Alvarez:2021zje} is essentially novelty detection with the score defined by the ratio of Poisson probabilities. CWoLa~\cite{Collins:2018epr, Collins:2019jip, Aad:2020cws}, TNT, SALAD and SULU are based on weakly supervised learning (for a review see Ref.~\cite{10.1093/nsr/nwx106}), where a DNN trained with a density-based algorithm was used as an evaluator. CwoLa was proposed for a resonance search, using the difference of signal fractions between the signal region and its sideband region. TNT and SALAD are the extensions, where TNT generalizes it to a broad-distributed signal search by assuming each event has at least two independent objects~\cite{Amram:2020ykb} and SALAD uses a reweighting method to eliminate the discrepancy between simulation and real data~\cite{Andreassen:2020nkr}. SULU shows a possibility to synthesize data and manages to provide unlabelled data points with soft labels~\cite{Dahbi:2020zjw}. UCluster~\cite{Mikuni:2020qds} constructs a graph network to define an embedded space and cluster collision events with similar properties in it. In this process, the locations of the testing events and cluster centroids in the embedded space are updated by considering the contribution of all testing events.
 
\begin{figure}[ht]
\centering
\begin{subfigure}[b]{0.3\textwidth} 
\includegraphics[width= \textwidth]{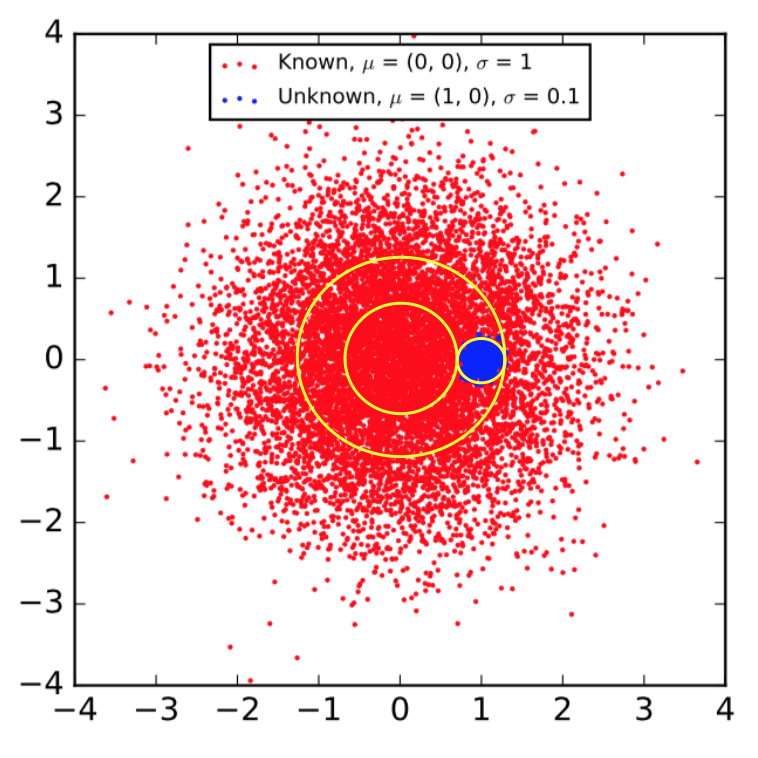}
\caption{}
\label{fig:a}
\end{subfigure}
\begin{subfigure}[b]{0.3\textwidth} 
\includegraphics[width=\textwidth]{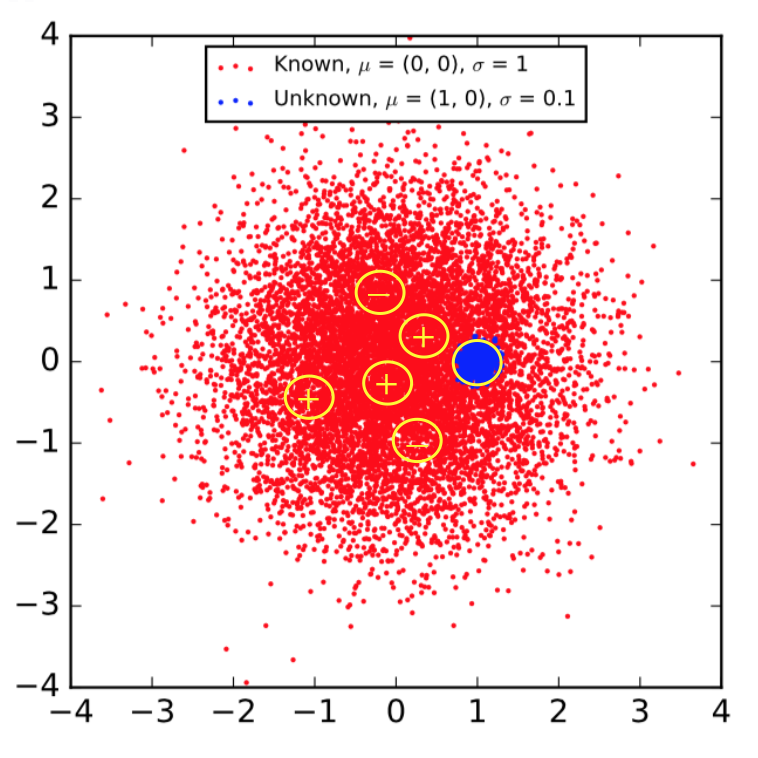}
\caption{}
\label{fig:b}
\end{subfigure}
\begin{subfigure}[b]{0.3\textwidth} 
\includegraphics[width= \textwidth]{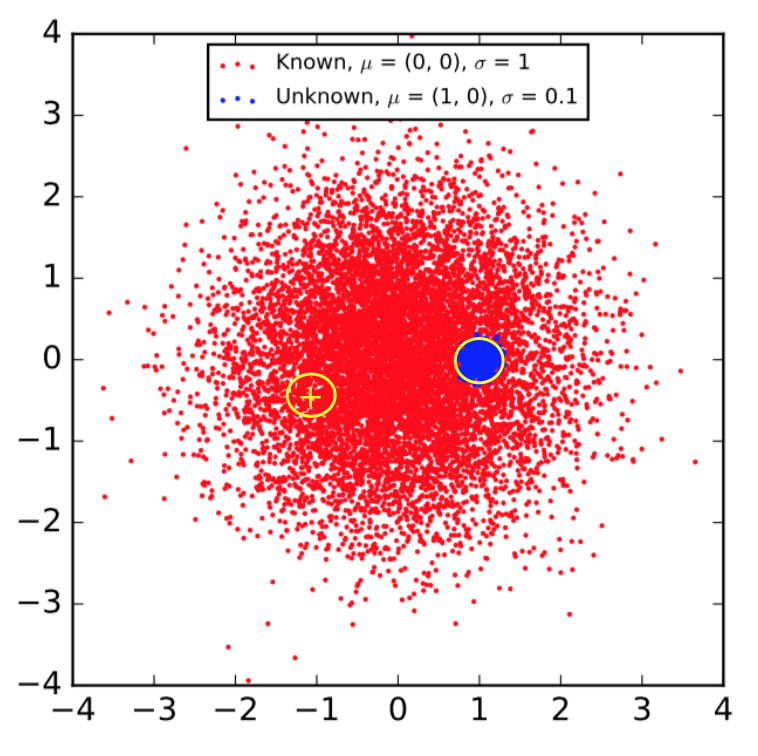}
\caption{}
\label{fig:c}
\end{subfigure}\\
\begin{subfigure}[b]{0.3\textwidth} 
\includegraphics[width=\textwidth]{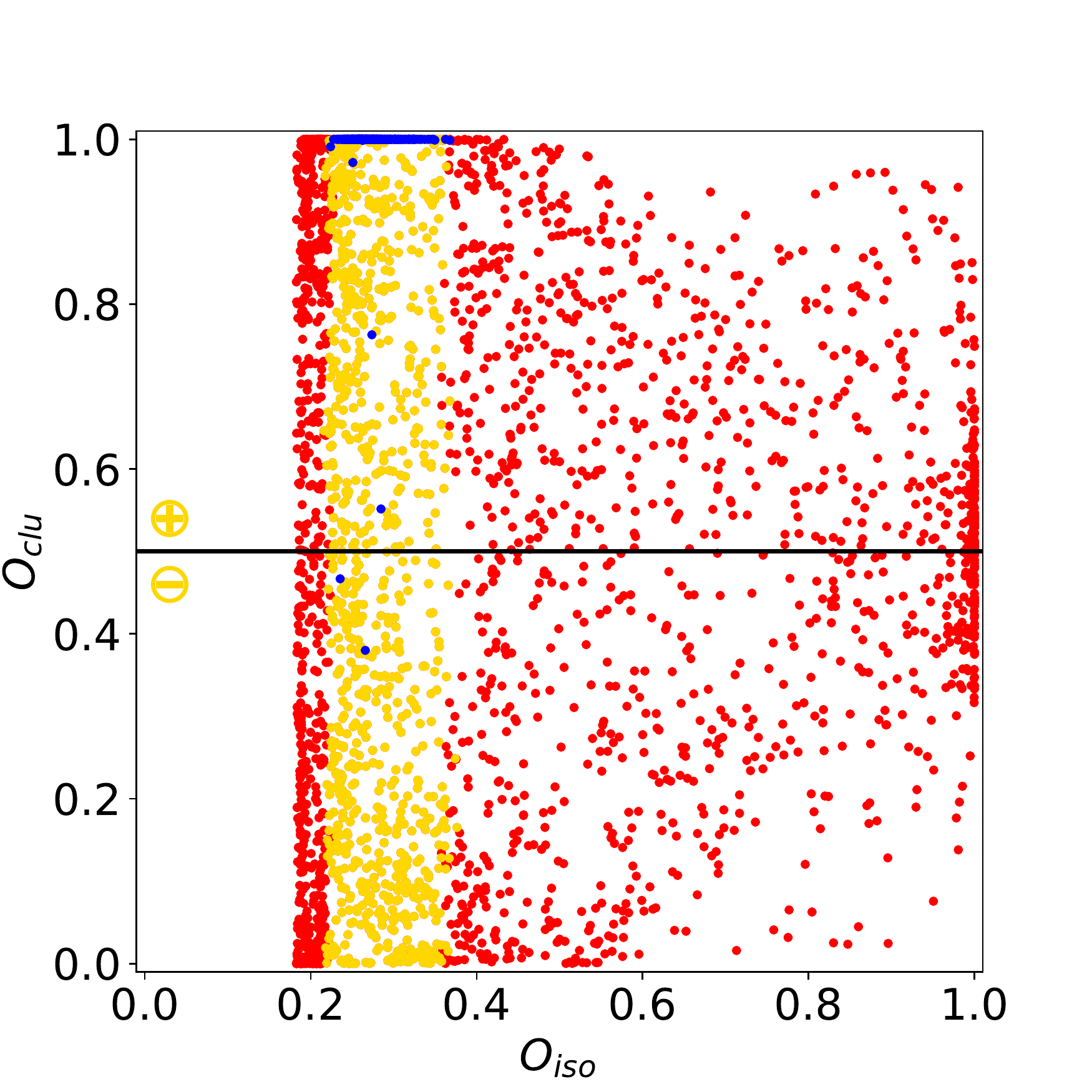}
\caption{}
\label{fig:d}
\end{subfigure}
\caption{Cartoon illustrating the response of different types of novelty evaluators in the case of a 2D Gaussian example, consisting of 10000 background events (red points) and 1000 signal events (blue points). In the case of ${\mathcal O_{\rm iso}}$ evaluators, the backgrounds are mainly from the ring band between two yellow circles in (a). In contrast, ${\mathcal O_{\rm clu}}$ evaluators are affected by the upward fluctuations in the local density of background events, denoted by the areas within the yellow-plus circles in (b) (downward fluctuations are denoted by the yellow-minus circles). A synergy-based evaluator, ${\mathcal O_{\rm syn}}$, would be affected only by the upward fluctuations  in local density that lay within the ring band, as illustrated in (c). The distribution of the ${\mathcal O_{\rm clu}}$ vs. ${\mathcal O_{\rm iso}}$ responses is shown in (d), where yellow points represent the background events within the yellow circle in (a), and the red points are the rest of background events. Exploiting the information encoded in both evaluators would clearly allow to improve the signal sensitivity by suppressing the backgrounds described above.}
\label{fig:syn}
\end{figure}  

Despite this great progress, challenges exist for these evaluators/algorithms that are closely related to their applicability:
\begin{enumerate}

\item In terms of signal events, the isolation-based evaluators ${\mathcal O_{\rm iso}}$ score high for events that are far from the bulk of the backgrounds in the feature space, such as those sitting on the tail of the distribution, while the clustering-based evaluators ${\mathcal O_{\rm clu}}$ tend to be sensitive for events that have a big deviation from the background-only prediction in their local density, such as those arising from resonance. In summary, these two types of evaluators attempt to detect novelty along two separate but special directions, which to some extent limits their respective signal coverage. 
 
\item In terms of backgrounds, the $\mathcal{O}_{\rm iso}$ evaluators may suffer from the presence of events that have comparable distance to the bulk of the backgrounds, while the $\mathcal{O}_{\rm clu}$ evaluators may suffer from those events that have a big fluctuation in their local density. These features are illustrated in Fig.~\ref{fig:syn} with a cartoon of 2D Gaussian samples. 

\end{enumerate}

\noindent 
These differences between $\mathcal{O}_{\rm iso}$ and $\mathcal{O}_{\rm clu}$ are rooted in the characters of the collider events employed to define $\mathcal{O}_{\rm iso}$ and $\mathcal{O}_{\rm clu}$, respectively. Given the lack of prior knowledge on the signal pattern and the relevant backgrounds, a synergistic treatment on the basis of these two types of evaluators is strongly required. This motivates the construction of a third type of novelty evaluator:    
\begin{itemize}

\item Synergy-based (${\mathcal O_{\rm syn}}$). The novelty response for a given testing event is evaluated according to a combination of its novelty responses to the isolation-based and clustering-based evaluators, $i.e.$ by a function $f ({\mathcal O_{\rm iso}},{\mathcal O_{\rm clu}})$. 

\end{itemize}
We expect that the synergy-based ${\mathcal O_{\rm syn}}$ evaluator can take advantage of the complementarity between the isolation-based and clustering-based evaluators and provide an option with much wider applicability.  Actually, essentially motivated by such a consideration, an evaluator has been proposed in~Ref. \cite{Hajer:2018kqm}, namely  
\begin{equation}
\label{eq:o_syn}
{\mathcal O_{\rm syn}} =\sqrt{{\mathcal O_{\rm iso}} {\mathcal O_{\rm clu}} }.
\end{equation}  

\noindent This design is not optimized, however. Firstly, the advantages of ${\mathcal O_{\rm iso}}$ are not fully exploited. As we will see, the signal events from the same region in feature space tend to have similar ${\mathcal O_{\rm iso}}$ scores. This fact essentially is a reflection of the signal-event ``clustering'' in feature space. But, the information of signal-event  clustering is not well-picked by ${\mathcal O_{\rm syn}}$ if the ${\mathcal O_{\rm iso}}$ score happens to be low. Secondly, even if ${\mathcal O_{\rm syn}}$ has a better performance than ${\mathcal O_{\rm iso}}$ and ${\mathcal O_{\rm clu}}$, there is a room to further improve the separation between the signal and the background events. Motivated by these considerations, we design an analysis scheme for novelty detection, where the complementarity between ${\mathcal O_{\rm iso}}$ and ${\mathcal O_{\rm clu}}$ could be fully utilized, in this paper. We will demonstrate that indeed this design yields a broad coverage of signal patterns and excellent property for generalization.   

This paper is organized as follows. In Sec.~\ref{sec:S2}, we will introduce the proposed analysis scheme for novelty detection and demonstrate its performance using the 2D Gaussian example, using the $k$NN designs for illustration. We will especially discuss in Subsec.~\ref{subsec:gen} the potential generalization of this scheme from the $k$NN-based evaluators to other ones listed in Tab.~\ref{tab:algorithms}. In Sec.~\ref{sec:S3}, this scheme will be applied to a more realistic use case, $i.e.$ the LHC detection of SM $t\bar t h$ production and of direct stop-quark pair production ($\tilde t \bar{\tilde t}$), both in the $t\bar{t}\gamma\gamma$ channel. Finally, we will conclude in Sec.~\ref{sec:S4}.

\section{A Complementarity-Based Analysis Scheme for Novelty Detection}
\label{sec:S2}

Below we will present an optimized analysis scheme for novelty evaluation at colliders and study its performance. Two classes of samples are relevant: the training (or reference) sample will (1) assist the evaluation of novelty response for a given event and (2) set up a reference on the novelty response of the known-pattern data, while the testing sample represents ``real'' data. To demonstrate the relevant points, we will use the $k$NN designs for the isolation-based and clustering-based evaluators proposed in Ref.~\cite{Hajer:2018kqm}, namely 
\begin{eqnarray}
\mathcal{O}_{\rm iso} = \frac{1}{2} \bigg( 1+ \text{erf} \frac{\Delta_{\rm iso}}{c\sqrt{2}}\bigg),~~~~ &\text{with}&~~~~\Delta_{\rm iso} = \frac{d_{\rm train} - \big< d'_{\rm train}\big>}{\big< d'_{\rm train}\big>^{1/2}} \ , \\
\mathcal{O}_{\rm clu} = \frac{1}{2} \bigg( 1+ \text{erf} \frac{\Delta_{\rm clu}}{c\sqrt{2}}\bigg),~~~~ &\text{with}&~~~~\Delta_{\rm clu} = \frac{d^{-m}_{\rm test} - d^{-m}_{\rm train}}{d^{-m/2}_{\rm train}} \ .  \label{eq:OO}
\end{eqnarray}
Here $d_{\rm train}$ is the mean distance of a testing event to its $k$ nearest neighbors in the training sample; $\big< d'_{\rm train}\big>$ is the reference mean distance defined by the training sample only;  $d_{\rm test}$ is the mean distance of the testing event to its $k$ nearest neighbors in the testing dataset; $m$ is the dimension of the latent space which we specify while performing a concrete analysis; and $c$ is a scaling factor chosen as the root mean square of the $\Delta_{\rm iso}$ ($\Delta_{\rm clu}$) scores for all testing data. For simplicity, the $k$NN distance metric is defined to be Euclidean. Please note: {\it this never means that the scheme developed here exclusively relies on the $k$NN designs for $\mathcal{O}_{\rm iso}$  and $\mathcal{O}_{\rm clu}$}. It represents a broad class of such designs instead. We will discuss its generalization in detail in Sec.~\ref{subsec:gen}.

\subsection{Design}
\label{subsec:design}

\begin{figure}[ht]
\makebox[\textwidth][c]{%
  \includegraphics[width=1.\textwidth]{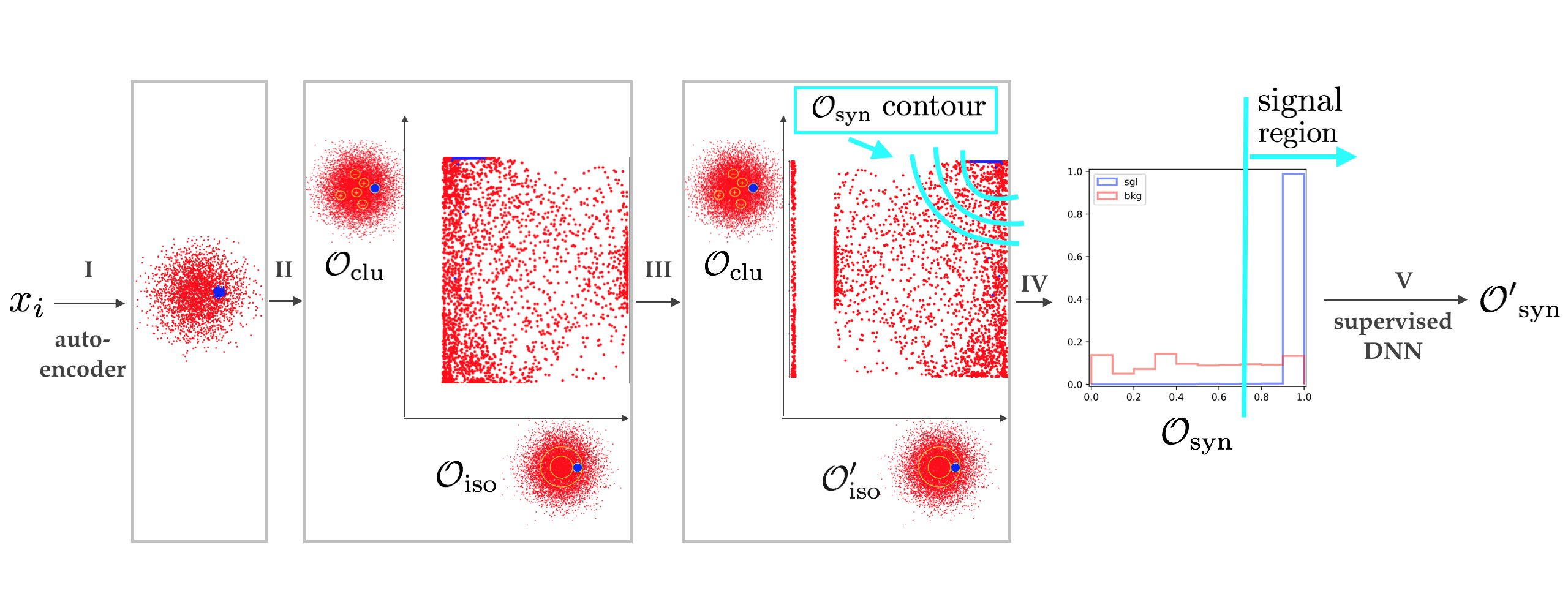}
}
\caption{Workflow for the proposed novelty-detection scheme at colliders.}
\label{fig:workflow}
\end{figure}  

The proposed analysis scheme for novelty detection at colliders involves five steps, which are outlined in~\fig{workflow} and described below:   
\begin{itemize}
\item Step I: dimensionality reduction. Reduce the high-dimensional feature space to a lower-dimensional latent space using a tool like AE~\cite{Vincent08} or its variances. This step can suppress the impact of statistical fluctuations in the high-dimensional feature space on detection efficiency, and generally allows a stronger response of anomalous data to the novelty evaluator.

\item Step II: novelty evaluation. Evaluate the novelty response of data, using the isolation-based and the clustering (density)-based methods, and project them to the ${\mathcal O_{\rm iso}} - {\mathcal O_{\rm clu}}$ plane for identifying the potential signal phase space. Note, with this step the data binned w.r.t. ${\mathcal O_{\rm clu}}$ may not respect the original statistics any more where the data are expected to be mutually independent. Although one could analyze potential signal events at this 2D plane directly (using, e.g., ${\mathcal O_{\rm syn}}$~\cite{Hajer:2018kqm}), it is possible to proceed with the following steps to attempt to build a more optimal discriminant. This is especially valuable for the cases where the backgrounds are rich and also widely distributed.

\item Step III: bin resorting of ${\mathcal O_{\rm iso}}$. The goal is to move the signal events to the top-right corner of the ${\mathcal O_{\rm iso}} - {\mathcal O_{\rm clu}}$ plane (recall, if the signal events are located in the bulk of the background distribution, they tend to be scored low by ${\mathcal O_{\rm iso}}$), such that they can be isolated with a cut based on the evaluator geometric mean (see Step IV). For this purpose, we define $n$ bins based on the ${\mathcal O_{\rm iso}}$ scores and then calculate $\xi_i= \frac{N_i- B_{\rm ref,i}}{\sqrt{B_{\rm ref,i}}}$ of the $i$-th bin. Here $N_i$ and $B_{\rm ref,i}$ are the event number of the $i$-th bin in the testing and training samples, respectively.  Assuming that $\xi_i$ is maximal for bin $i'$, we propose a moving strategy as follows: 
\begin{itemize} 
\item $i' > n/2$. We will shift all bins to the right so that bin $i'$ ($i' +1$) becomes the most right bin if $\xi_{i'+1}< 1$ ($\xi_{i'+1} \geq 1$). 
\item $i' \leq n/2$. We will shift all bins to the left so that bin $i'$ ($i' - 1$) becomes the most left bin if $\xi_{i'-1} < 1$ ($\xi_{i'-1} \geq 1$), and then assign a new score $1 - {\mathcal O_{\rm iso}}$ to each event.

\end{itemize}
This strategy preserves high priority for the bin with a large $ \xi_i$ value. For the convenience of discussions, below we define the new score after movement as ${\mathcal O'_{\rm iso}}$. We will use 10 uniform bins for the resorting to get ${\mathcal O'_{\rm iso}}$.   
 
\item Step IV: signal-like region identification. We define the signal-like region to be  
\begin{equation}
S': \quad O_{\rm syn} = \sqrt{{\mathcal O'_{\rm iso}}  {\mathcal O_{\rm clu}}} > r_0\ .   \label{eq:Opsyn}
\end{equation}
Here $r_0$ is a threshold. It defines the boundary of the signal-like region in the $O'_{\rm iso} - O_{\rm clu}$ plane. The determination of $r_0$ is not unique. For example, we can choose it to be the one above which we have the largest $\frac{N- B_{\rm ref}}{\sqrt{B_{\rm ref}}}$, with $N$ and  $B_{\rm ref}$ being the numbers of the testing and simulated background events in this region, respectively. In this study we will use a threshold $r_0 = 0.7$ or $0.6$, unless otherwise specified. Alternatively, one can identify the signal-like region by collecting the 2D bins with $\frac{N-B_{\rm ref}}{\sqrt{B_{\rm ref}}} > r_0$ at the ${\mathcal O_{\rm iso}} - {\mathcal O_{\rm clu}}$ plane. In the case, $S'$ tends to be contaminated by fluctuating non-signal patches given the finite size of the samples. 

\item Step V: novelty re-evaluation. Construct a DNN of weakly supervised learning, to distinguish the data falling in the $S'$ region and the simulated background training samples. The inputs can be the full set of kinematic features or only the ones defining the latent space. The novelty of all testing data then will be re-evaluated by a new evaluator, $i.e.$, the output neuron (denoted as $\mathcal{O'}_{\rm syn}$). In this paper, we build this network with a simple architecture (five layers with 2, 8, 6, 4, 1 neurons, respectively), with the input neurons being the latent-space dimensions (see more discussions below). To ensure its robustness, we could train this model with multiple initial seeds and average all $\mathcal{O'}_{\rm syn}$ scores for each data point. We will train twenty such classifiers in total to define $\langle \mathcal{O'}_{\rm syn} \rangle$, namely the average of $\mathcal{O'}_{\rm syn}$, for all scenarios in this study. Different from the cases of $\mathcal{O}_{clu}$ and $\mathcal{O}_{syn}$, the data binned by $\mathcal{O'}_{syn}$ respect the Gaussian/Poisson statistics. As to be seen, the application of $\mathcal{O'}_{\rm syn}$ strengthens the effectiveness and efficiency of novelty evaluation, and finally optimizes the sensitivity of the detection. All neural networks used in this study will be trained with Keras~\cite{chollet2015keras}. 

\end{itemize}
After Steps I-V, we will be able to calculate the sensitivity reach in this analysis scheme. Explicitly, we will use the Poisson-statistics-based formula  
\begin{eqnarray}
Z = \sqrt{2(B+S) \log{\frac{B+S}{B}} - 2 S}
\end{eqnarray}
to calculate the significance of excluding the background-only hypothesis with $\mathcal{O'}_{\rm syn}$.

\subsection{Performance}
\label{subsec:performance}

\begin{figure}[ht]
\centering\includegraphics[width=0.9 \textwidth]{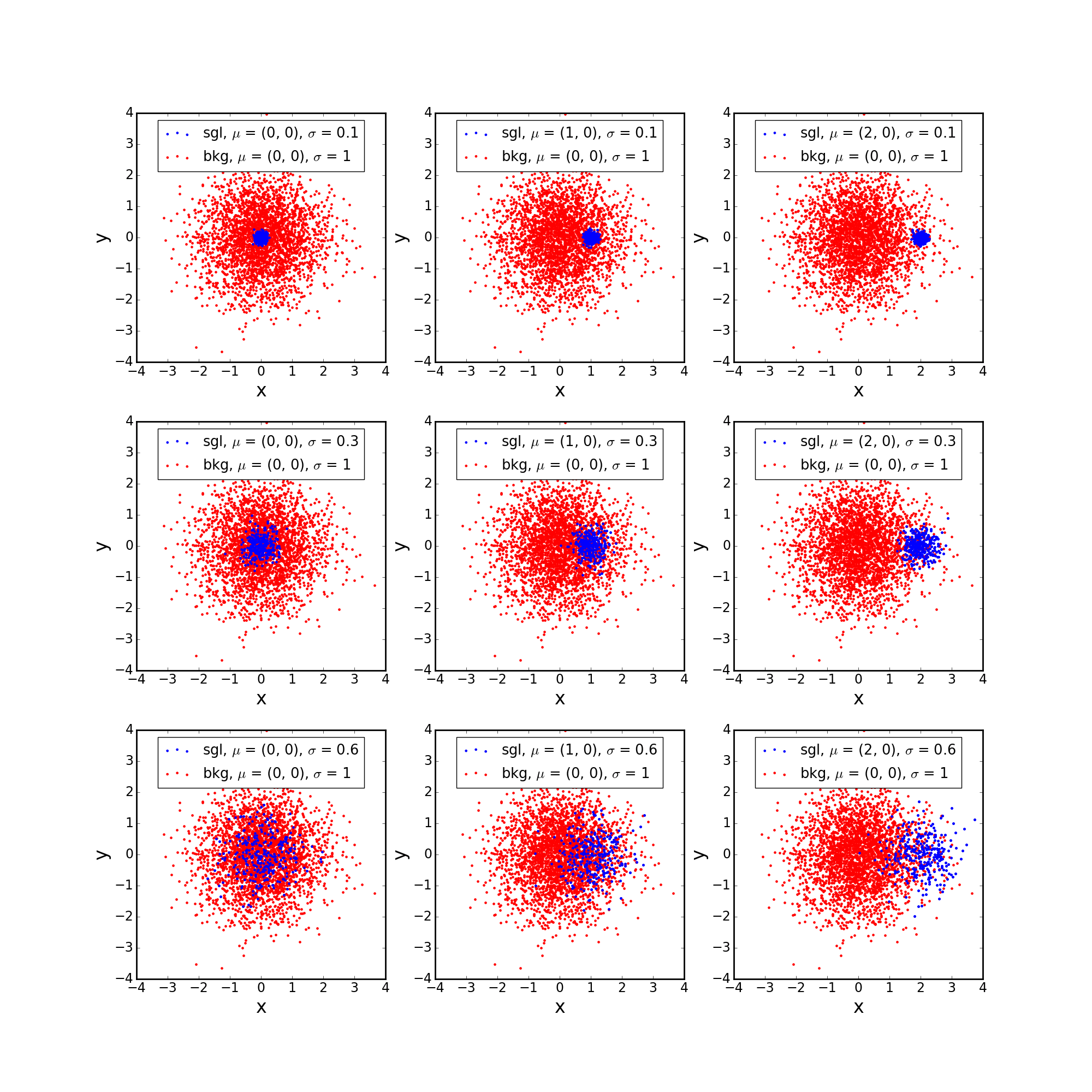}
\caption{2D Gaussian samples used for testing. Each contains 100000 background events and 1000 signal events. For convenience, we label these benchmark points, as ``BP 1'', ... , ``BP 9'',  from left to right and from top to bottom.}
\label{fig:gaussian}
\end{figure}  

Below we will explore the sensitivity performance of the proposed working scheme for novelty detection, with the signals and background events in the latent space being mimicked by 2D Gaussian samples.\footnote{In a realistic analysis, the latent space after dimensionality reduction could have a dimension higher than two. Its definition is essentially determined by the complexity of the potential signals to detect.} Concretely, the backgrounds are defined as $\mathcal N((0,0), \mathbf I)$, $i.e.$ a Gaussian sample centered at the origin with a standard deviation equal to unity, with the signals $\mathcal N((\mu,0), \sigma^2 \mathbf I)$ sitting on its top. In total, nine different signal benchmarks (denoted "benchmark points, or BPs) are considered, with $\mu = 0$, $1$, $2$ and $\sigma= 0.1$, $0.3$, $0.6$, all of which are unknown to the detection. The distributions of these signals, together with that for the backgrounds, are shown in \fig{gaussian}, where $x$ and $y$ represent two directions defining the latent space.
As $\mu$ increases, the signal distribution moves along the $x$ axis from the background center to its tail. As $\sigma$ increases, the signal distribution evolves from a narrow peak to a broad distribution. 

For this study we generate $4\times 10^5$ (background) events for training, and $1\times 10^3 + 1\times 10^5$ (signal + background) events for testing. For the ${\mathcal O_{\rm iso}}$ evaluator, $d_\text{train}$ and $\langle d^\prime_\text{train} \rangle$ are calculated with $k=4000$, and for the ${\mathcal O_{\rm clu}}$ evaluator,  $d_\text{train}$ and $d_\text{test}$ are calculated with $k=4000$ and $k=1000$, respectively. The corresponding scores can be found in App.~\ref{sec:AA}. Notably, despite the impact on ${\mathcal O_{\rm iso}}$ and ${\mathcal O_{\rm clu}}$ (and hence on ${\mathcal O_{\rm syn}}$) when varying the $k$ value from $\ll N_{\rm sig}$ to $\sim N_{\rm sig}$, we find that the ${\mathcal O'_{\rm syn}}$ performance is robust against this variation.

\begin{figure}[ht]
\centering\includegraphics[width=0.9 \textwidth]{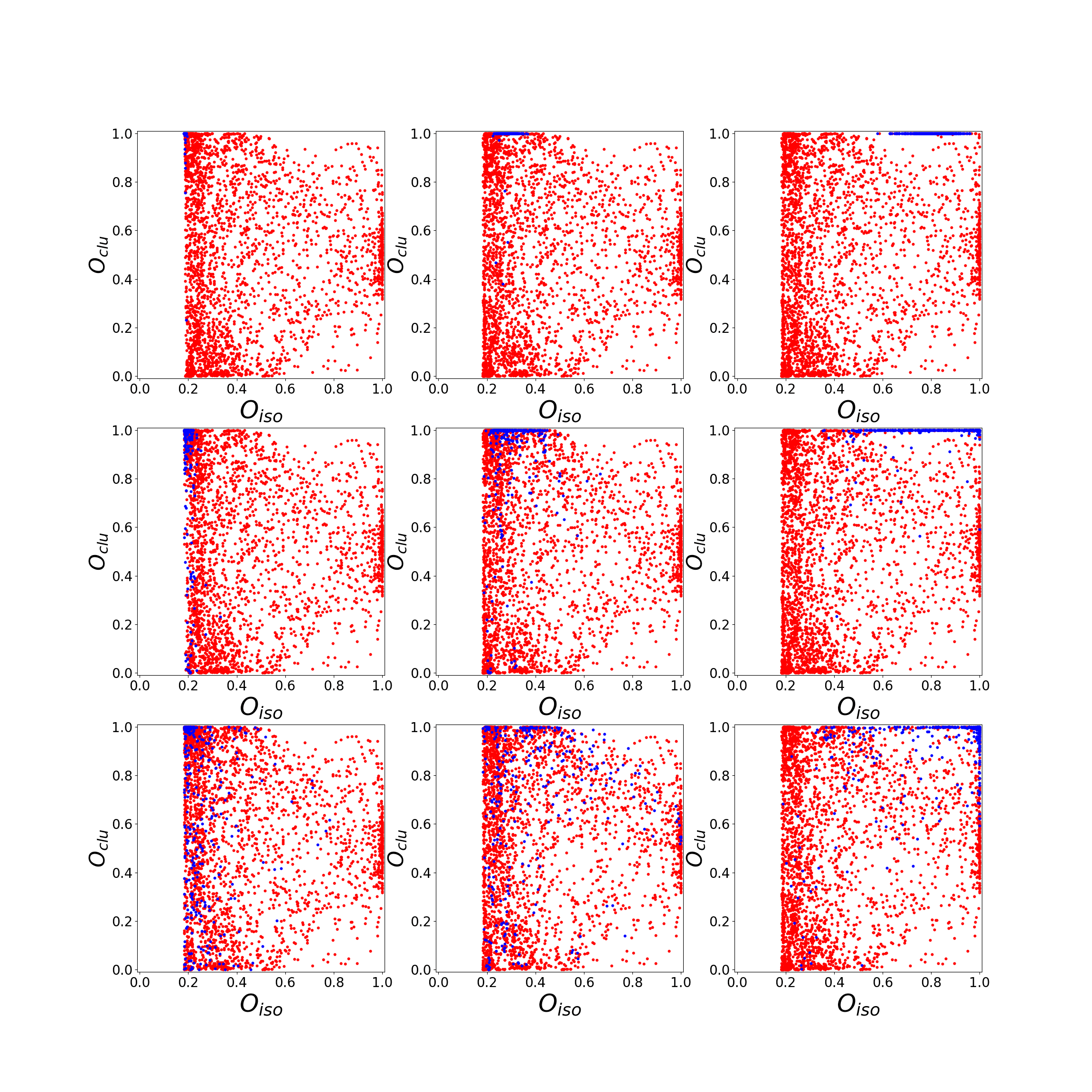} 
\caption{Distribution of novelty scores in the $\mathcal O_{\rm iso}-\mathcal O_{\rm clu}$ plane for the 2D Gaussian testing samples corresponding to the same BPs as in~\fig{gaussian}.}
\label{fig:gaussscore}
\end{figure}  

\begin{figure}[ht]
\centering\includegraphics[width=0.9 \textwidth]{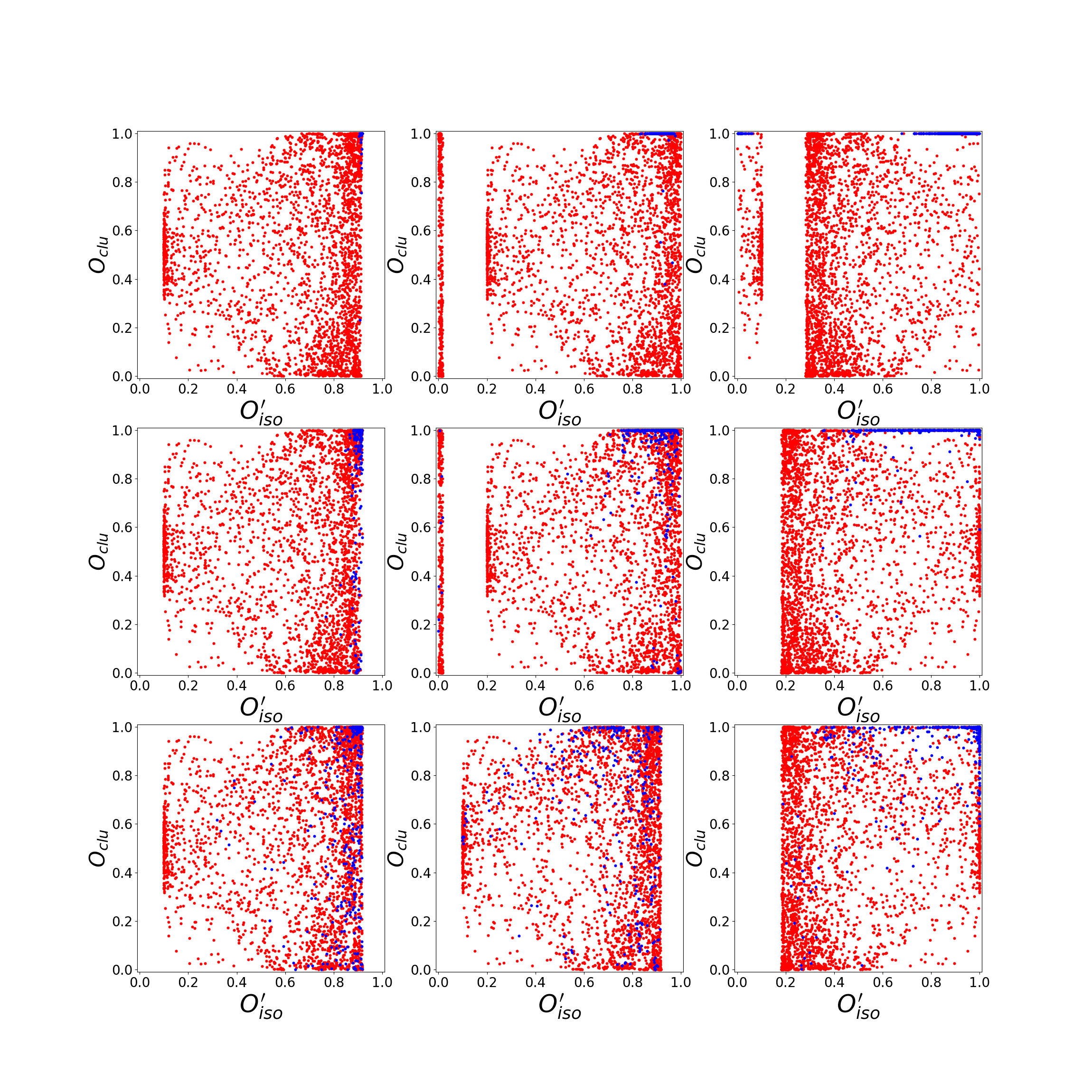}
\caption{Distribution of novelty scores in the $\mathcal O'_{\rm iso}-\mathcal O_{\rm clu}$ plane for the 2D Gaussian testing samples corresponding to the same BPs as in~\fig{gaussian}.} 
\label{fig:gaussscoreafter}
\end{figure}  

As an isolation-based evaluator, the performance of ${\mathcal O_{\rm iso}}$ is highly sensitive to the $\mu$ parameter. As shown in~\fig{gaussscore}, the signal events tend to score lower as the $\mu$ value decreases. This is the result of the ${\mathcal O_{\rm iso}}$ definition, since for lower $\mu$ values signal events are located in the bulk of the background distribution and $d_{\rm train}$ becomes numerically closer to $\big< d'_{\rm train}\big>$. The performance of ${\mathcal O_{\rm iso}}$ for the different BPs is illustrated in~\fig{AUC_ROC}, which summarizes the ROC curves and AUC values. As expected, the BPs in the left column have a smaller AUC value for ${\mathcal O_{\rm iso}}$ than their respective counterparts in the right column. Compared to variations in the $\mu$ parameter, the impact of the $\sigma$ parameter on the ${\mathcal O_{\rm iso}}$ performance is relatively weak. It should be noted that the degraded novelty response in the small-$\mu$ BPs does not imply that the signal events are not well-picked by ${\mathcal O_{\rm iso}}$. It just means that the score in such a case is not a proper measure of the novelty. Actually, from~\fig{gaussscore} one can see that the signal events are still clustered in some specific ${\mathcal O_{\rm iso}}$ bins, despite their low scores in the small-$\mu$ BPs. This reveals one simple but important fact for both isolation-based and clustering-based evaluators: {\it signal events from the same region in feature space tend to be scored the same}. In this analysis scheme, we take a bin-resorting for ${\mathcal O_{\rm iso}}$ (we redefine it as ${\mathcal O'_{\rm iso}}$ for the convenience of presentation), and demonstrate the outcomes after bin resorting in Fig.~\ref{fig:gaussscoreafter} and the correspondent ROC curves and AUC values in \fig{AUC_ROC}. As expected, the signal events tend to be scored high by ${\mathcal O'_{\rm iso}}$, which eventually yields an AUC value $\ge 0.5$. 

As a clustering-based evaluator, $O_{\rm clu}$ performs excellently in detecting a resonance-like structure. As shown in  \fig{gaussscore} and Fig.~\ref{fig:gaussscoreafter}, most signal events are scored very high for the BPs with $\sigma = 0.1$. Yet, as expected, the performance of $O_{\rm clu}$ degrades as $\sigma$ increases, since for a broader signal distribution it becomes increasingly harder to detect event clustering. The same trend can be also seen from the ROC curves of $O_{\rm clu}$ and their AUC values shown in \fig{AUC_ROC}. Indeed, the obtained AUC values are all close to one for the BPs with $\sigma=0.1$, but they are reduced as $\sigma$ becomes larger. Compared to the $\sigma$ parameter, the impact of the $\mu$ parameter on the $O_{\rm clu}$ performance is relatively weak. But still, as $\mu$ increases, the obtained AUC values gradually improve, due to the reduced overlap between the signal and background distributions. 

\begin{figure}[t]
\centering\includegraphics[width=1.0 \textwidth]{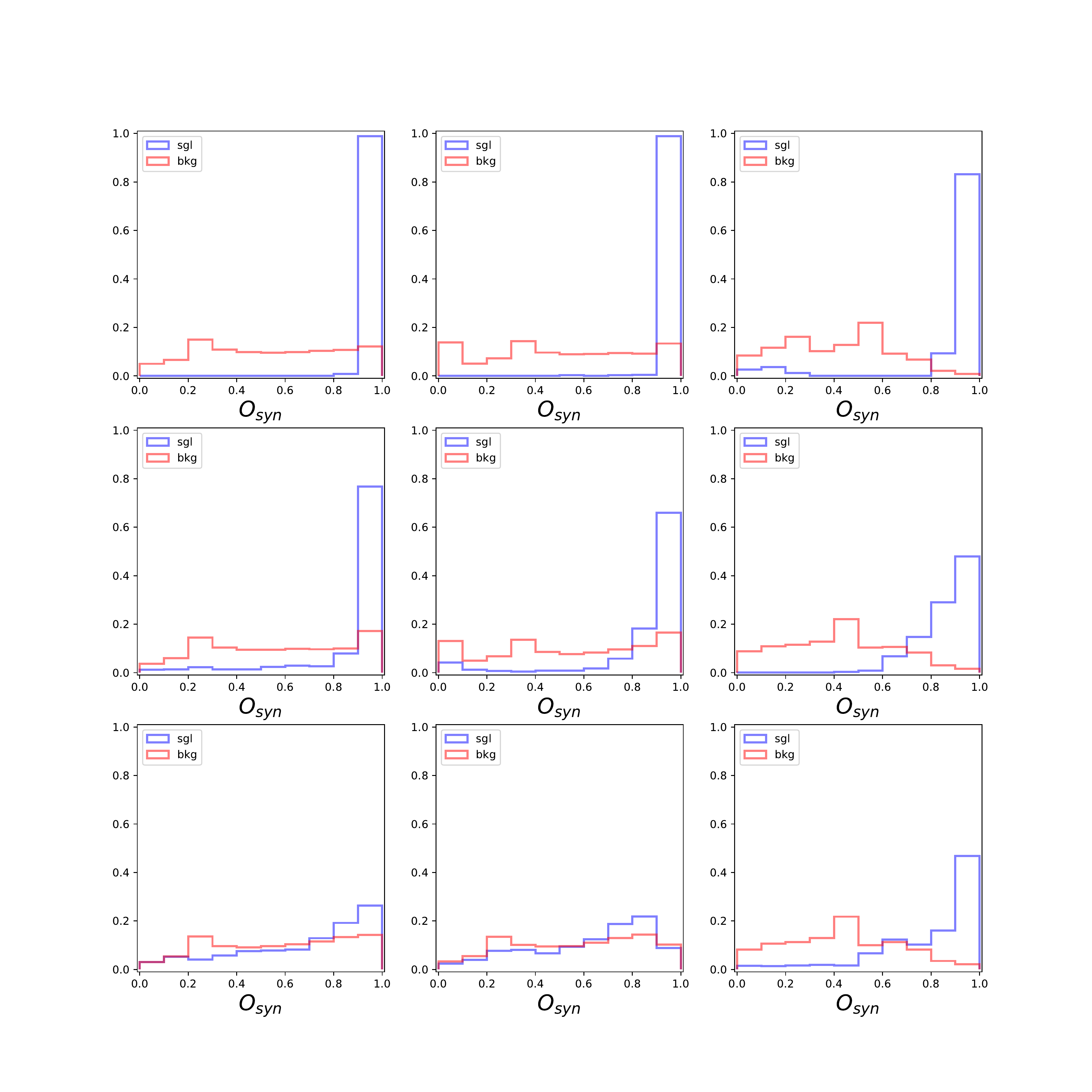}
\caption{Novelty score according to $O_{\rm syn}$ for the 2D Gaussian testing samples corresponding to the same BPs as in~\fig{gaussian}.}
\label{fig:osyn}
\end{figure} 

\begin{figure}[ht]
\centering\includegraphics[width=1.0 \textwidth]{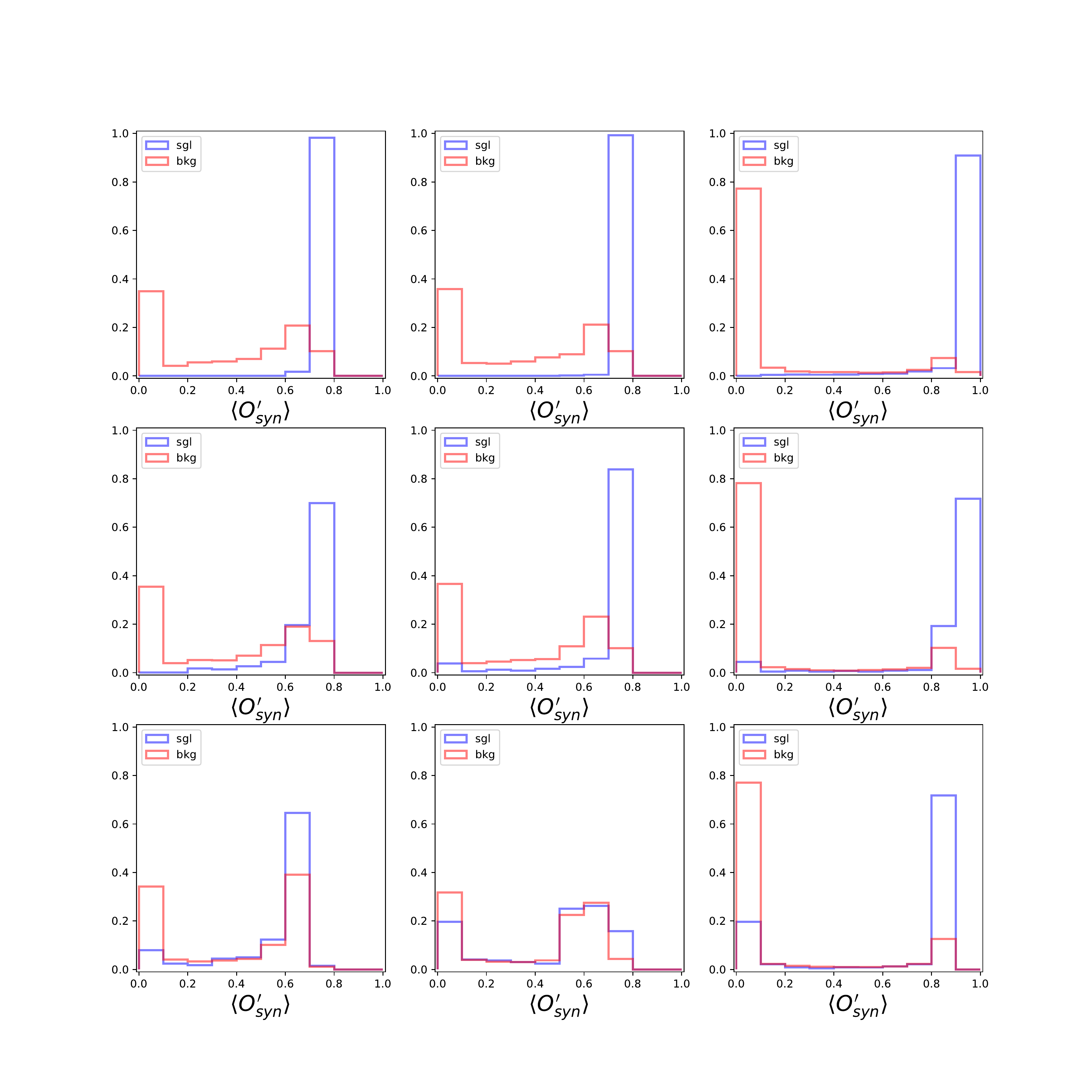}
\caption{Novelty score according to $\langle \mathcal O'_{\rm syn} \rangle$ for the 2D Gaussian testing samples corresponding to the same BPs as in~\fig{gaussian}.}
\label{fig:osyn_prime}
\end{figure} 

\begin{figure}[ht]
\centering\includegraphics[width=1.0 \textwidth]{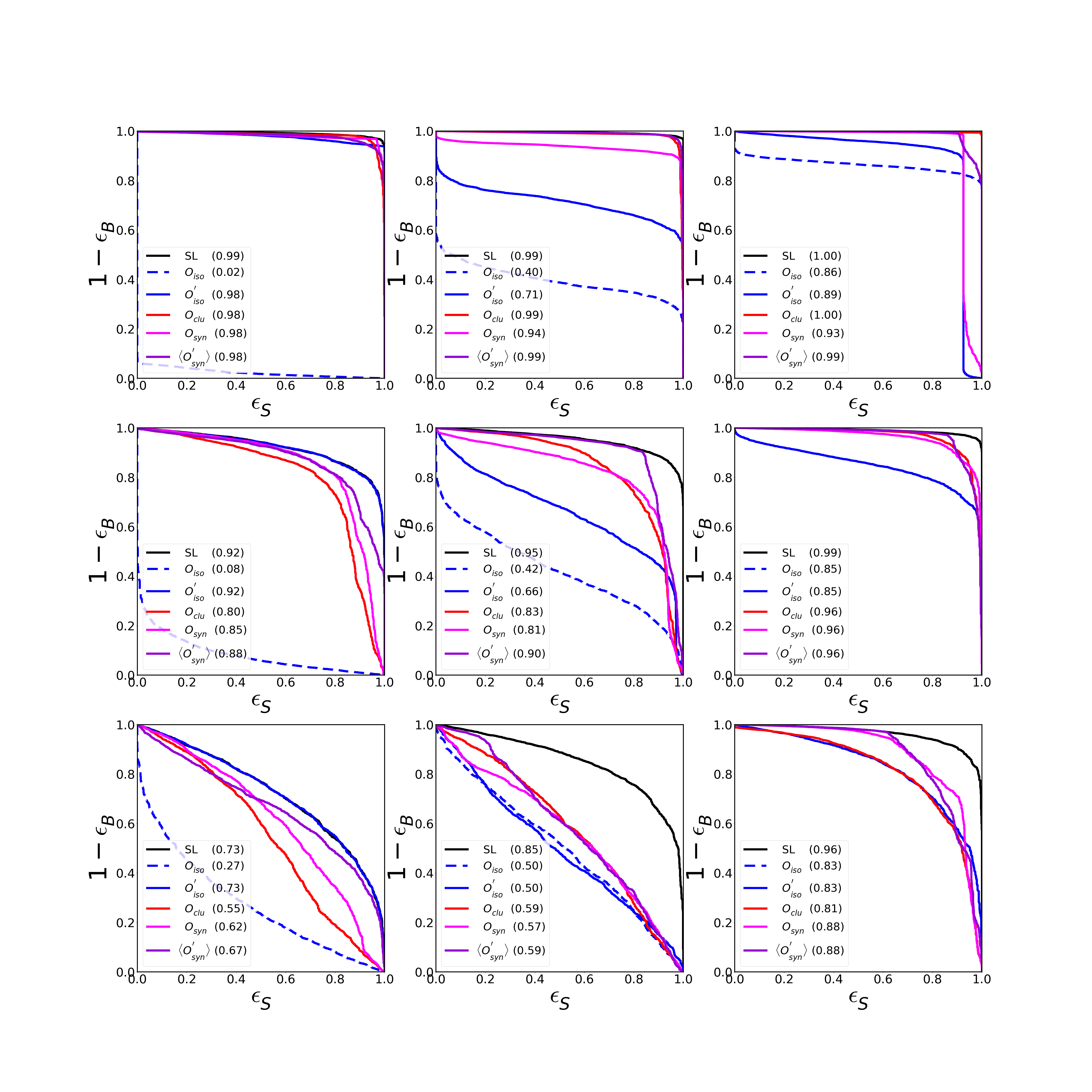}
\caption{ROC curves and their AUC values (given in brackets) for the set of novelty evaluators considered, applied to the 2D Gaussian testing samples corresponding to the same BPs as in~\fig{gaussian}. Here $r_0=0.7$ is taken to define the signal-like sample $S'$ for training ${\mathcal O'_{\rm syn}}$. As a reference, the performance from supervised learning (SL) has also been reported. The dashed and solid blue curves are fully overlapped in the panels of the right column.}
\label{fig:AUC_ROC}
\end{figure}

\begin{figure}[ht]
\centering\includegraphics[width=1.0 \textwidth]{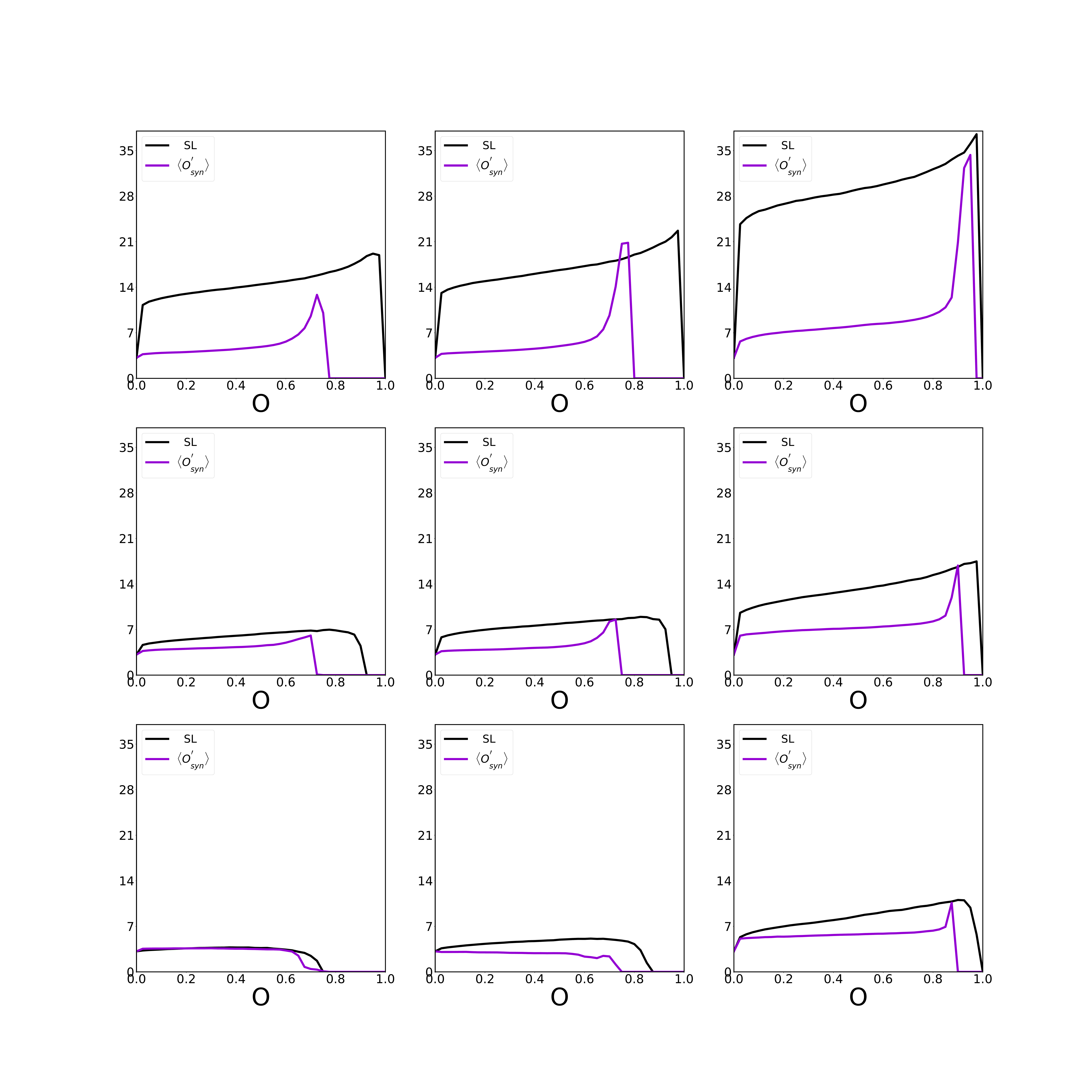}
\caption{Statistical significance as a function of the $\langle \mathcal{O'}_{syn} \rangle$ threshold (40 bins are defined), applied to the 2D Gaussian testing samples corresponding to the same BPs as in~\fig{gaussian}. Here $r_0=0.7$ is taken to define the signal-like sample $S'$ for training ${\mathcal O'_{\rm syn}}$. As a reference, the SL performance has also been reported.}
\label{fig:sigfg}
\end{figure}  

Following the previous discussions, the complementarity between the isolation-based and the clustering-based novelty evaluators is two-fold. First, $O_{\rm iso}$ (and ${\mathcal O'_{\rm iso}}$) is more sensitive to $\mu$ while $O_{\rm clu}$ is more sensitive to $\sigma$. This makes the BPs with small or large $\mu$ and small $\sigma$ relatively easy to probe, while leaves the BPs with intermediate $\mu$ and large $\sigma$ (e.g., BP 8) most difficult to detect. Second, the signal bins of $O_{\rm iso}$ and $O_{\rm clu}$ tend to be contaminated by the background events from different non-signal-like regions at the $x-y$ plane. Indeed, from Fig.~\ref{fig:gaussscoreafter} one can see that the signal events tend to cluster in the right-upper corner of the $O'_{\rm iso} - O_{\rm clu}$ plane, while many backgrounds with a high score of $O'_{\rm iso}$ and $O_{\rm clu}$ are scattered along the $O_{\rm clu}$ and  $O'_{\rm iso}$ directions, respectively. This provides an opportunity to define a signal bin region where those reducible backgrounds have been suppressed, by properly combining these two evaluators as, e.g., $\mathcal{O}_{\rm syn}$ in Eq.~(\ref{eq:Opsyn}). This is Step IV in this analysis scheme.  

The novelty responses of the 2D Gaussian samples to $O_{\rm syn}$, $\langle O'_{\rm syn} \rangle$ and their ROC curves are demonstrated in Fig.~\ref{fig:osyn}, Fig.~\ref{fig:osyn_prime} and \fig{AUC_ROC}, respectively. The significance curves based on $\langle O'_{\rm syn} \rangle$ are illustrated in Fig.~\ref{fig:sigfg}. By comparing the AUC values of the $\mathcal{O}_{\rm syn}$ ROC curves with those of ${\mathcal O'_{\rm iso}}$ and ${\mathcal O_{\rm clu}}$, one can see that this intuitive design performs universally better than at least one of ${\mathcal O'_{\rm iso}}$ and ${\mathcal O_{\rm clu}}$. Indeed, to some extent the complementarity discussed above has been picked up by $\mathcal{O}_{\rm syn}$.  The separation between the signal and background events gets further enhanced by ${\mathcal O'_{\rm syn}}$. 
Except for BPs 4 and 7, where ${\mathcal O'_{\rm iso}}$ performs almost perfectly, ${\mathcal O'_{\rm syn}}$ performs better or equally well for all BPs,   compared to the  other evaluators. This displays the broad applicability of ${\mathcal O'_{\rm syn}}$ for novelty detection.  
At last, we would point out: the AUC values and maximal significances based on  ${\mathcal O'_{\rm syn}}$ are generally close to or not far from their respective ``ideal'' values set by the dedicated supervised learning (except for BP 8). Here the SL models have been trained using the simulated signal and background samples directly~\footnote{For 2D Gaussian samples, the optimal discriminant values can also be computed analytically based on Neyman-Pearson lemma. We have checked that these analytical results are  comparable with those obtained using the supervised learning network.}.

\subsection{Generalization}
\label{subsec:gen}

As previously indicated, the proposed analysis scheme is general. One can pair any of the isolation-based and clustering-based evaluators listed in Table~\ref{tab:algorithms} and use them to replace the $k$NN-based  $\mathcal O_{\rm iso}$ and $\mathcal O_{\rm clu}$ used in this study, with the expectation of similar outcomes.\footnote{The complementarity between the AE reconstruction error and CWoLa was recently shown in Ref.~\cite{Collins:2021nxn}. In Ref.~\cite{Caron:2021wmq} it was suggested to detect anomalies by combining isolation-based Deep SVDD and clustering-based autoregressive flows. Besides, it was also pointed out in Refs.~\cite{Dillon:2021nxw} and~\cite{Finke:2021sdf} that the AE reconstruction error detects novelty only in one direction, as discussed in this paper. In relation to that, latent space tagging (a clustering-based method in our definition) was introduced to further improve the detection performance in Ref.~\cite{Dillon:2021nxw}. Essentially, these suggested methods exploit the complementarity between the isolation-based and clustering-based evaluators/algorythms, as was originally done for the $k$NN-based $\mathcal O_{\rm iso}$ and $\mathcal O_{\rm clu}$ in Ref.~\cite{Hajer:2018kqm}.} Alternatively, one can develop a clustering-based ``partner'' evaluator for each isolation-based evaluator, as it occurs to the $k$NN-based designs, and then apply them to this scheme. Recall, the definition of both $\mathcal O_{\rm iso}$ and $\mathcal O_{\rm clu}$ relies on some distance measure ``$d$'' (see Eq.~(\ref{eq:OO})). Here we only need to replace the $k$NN-based measure with the one defining the given isolation-based evaluator.  

To demonstrate these points (especially the second one), let us consider an AE-based novelty evaluator as an example. In particular, the AE reconstruction error ($R_{\rm AE}$) is essentially a distance measure in the feature space or an isolation-based evaluator. So we can define the $R_{\rm AE}$-based $\mathcal O_{\rm iso}$ and $\mathcal O_{\rm clu}$ by taking $d = R_{\rm AE}^{1/2}$ (the power of $\frac{1}{2}$ arises from the fact that $R_{\rm AE}$ is formally a sum of ``length'' square in the feature space). Here two AEs with the same architecture (denoted as AE-A and AE-B) need to be trained, using the training and testing samples respectively. Then $d_{\rm train}$ and $d'_{\rm train}$ will be calculated by AE-A, while $d_{\rm test}$ will be calculated by AE-B.     

As a concrete example, we apply this variant of our designed scheme to the analysis of the 2D Gaussian samples corresponding to BP 5. Here AE-A and AE-B are constructed with 3 hidden layers, which contain 2, 1 and 2 neurons respectively. Tanh is selected as the    activation function for all layers except the last one, where a linear function is applied to match the input and output ranges. Stochastic gradient descent is used as the optimizer, with its learning rate, decay rate and momentum being 0.3, 0.0001 and 0.99 respectively. The epoch number is fixed to 400. The AE-A is trained with a training sample of 400000 events, while AE-B is trained with a testing sample of 101000 events. To improve the analysis stability, we train both AE-A and AE-B with 20 random initial seeds and use the square root of the averaged reconstruction errors for each testing data, $i.e.$ $\langle R_{\rm AE} \rangle^{1/2}$, 
to define the distance ``$d$'' and eventually the evaluator $\mathcal O_{\rm syn}$. 

\begin{figure}[ht]
\centering
\begin{subfigure}[b]{0.32\textwidth} 
\includegraphics[width=\textwidth]{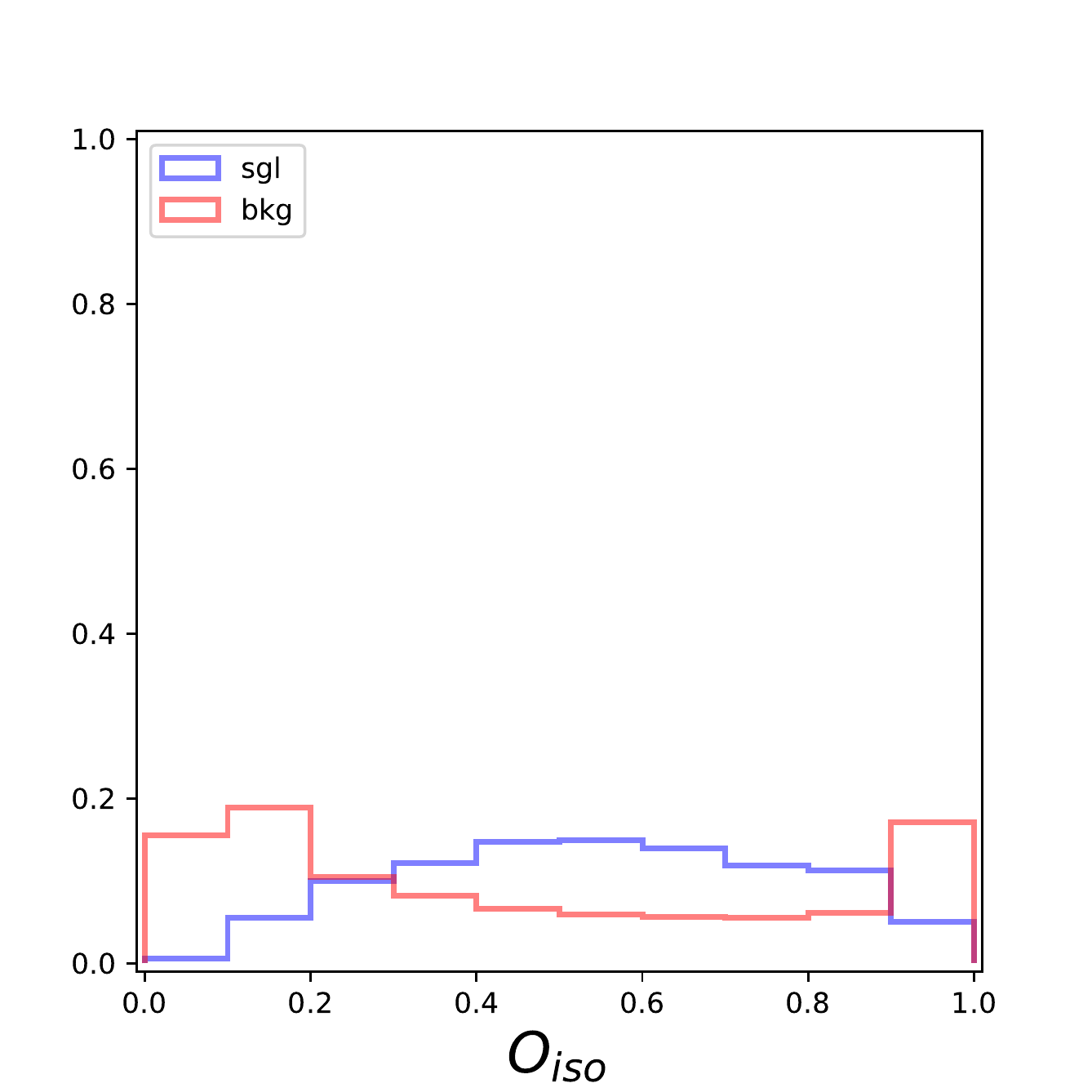}
\caption{}
\end{subfigure}
\begin{subfigure}[b]{0.32\textwidth} 
\includegraphics[width=\textwidth]{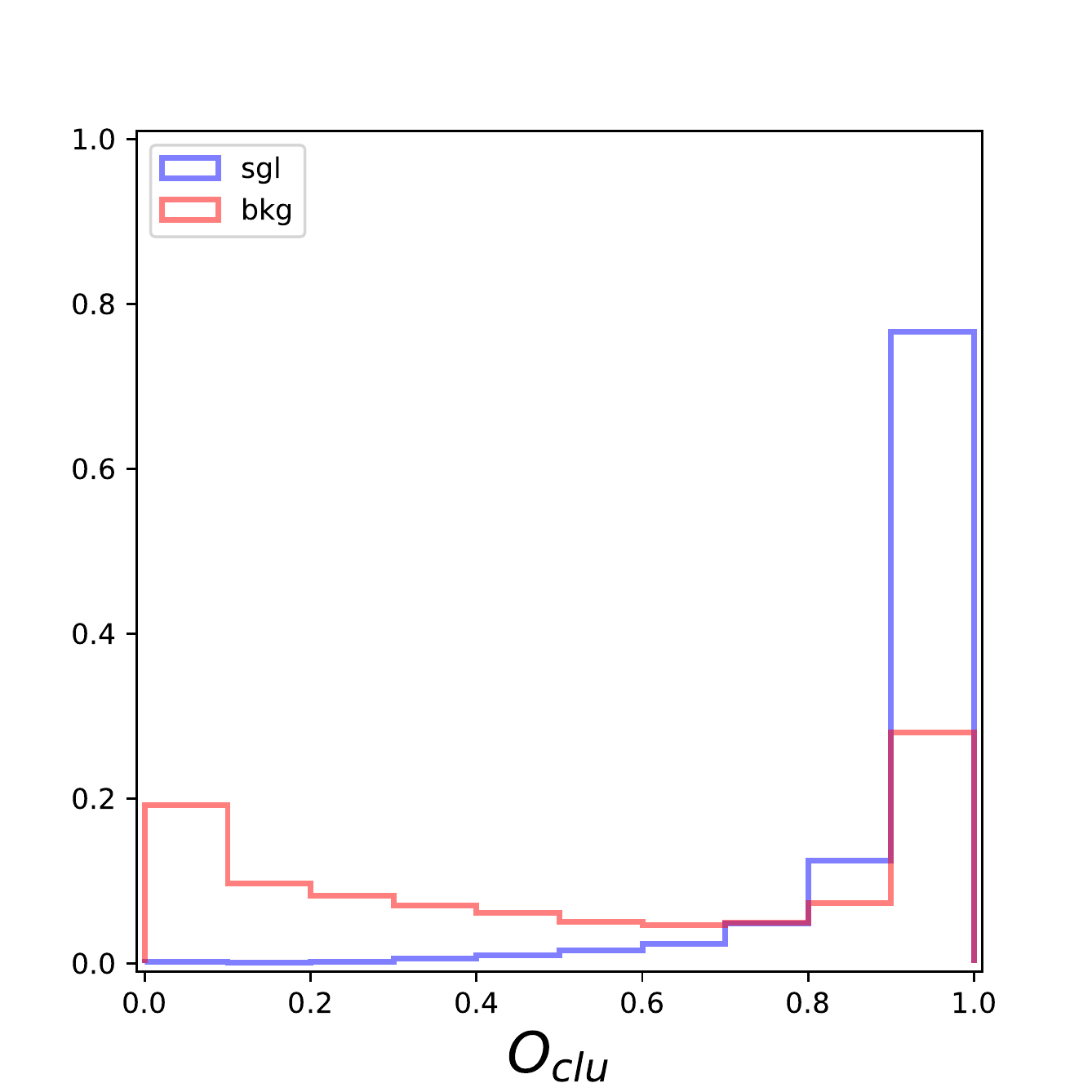}
\caption{}
\end{subfigure}
\begin{subfigure}[b]{0.32\textwidth} 
\includegraphics[width=\textwidth]{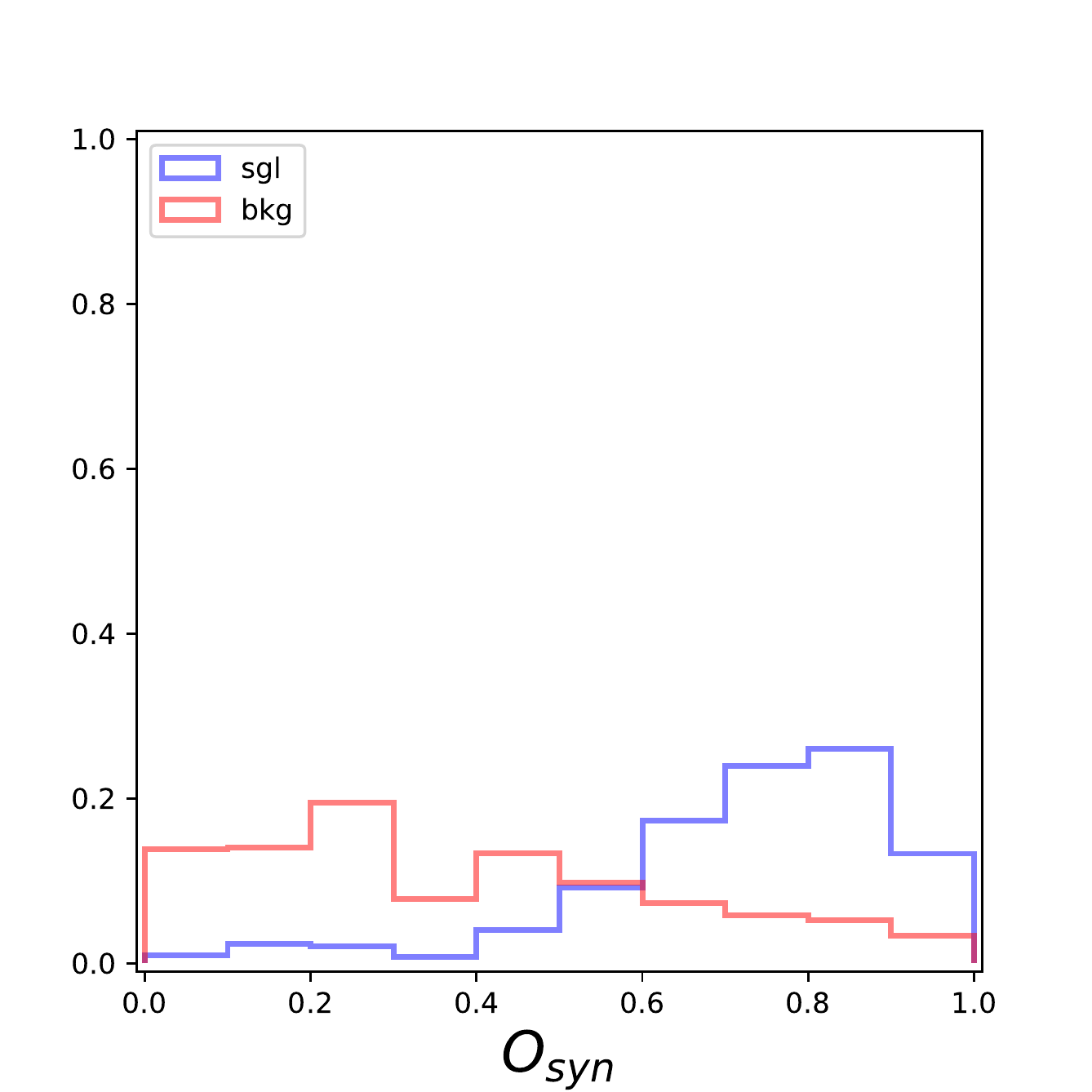}
\caption{}
\end{subfigure}
\begin{subfigure}[b]{0.32\textwidth} 
\includegraphics[width=\textwidth]{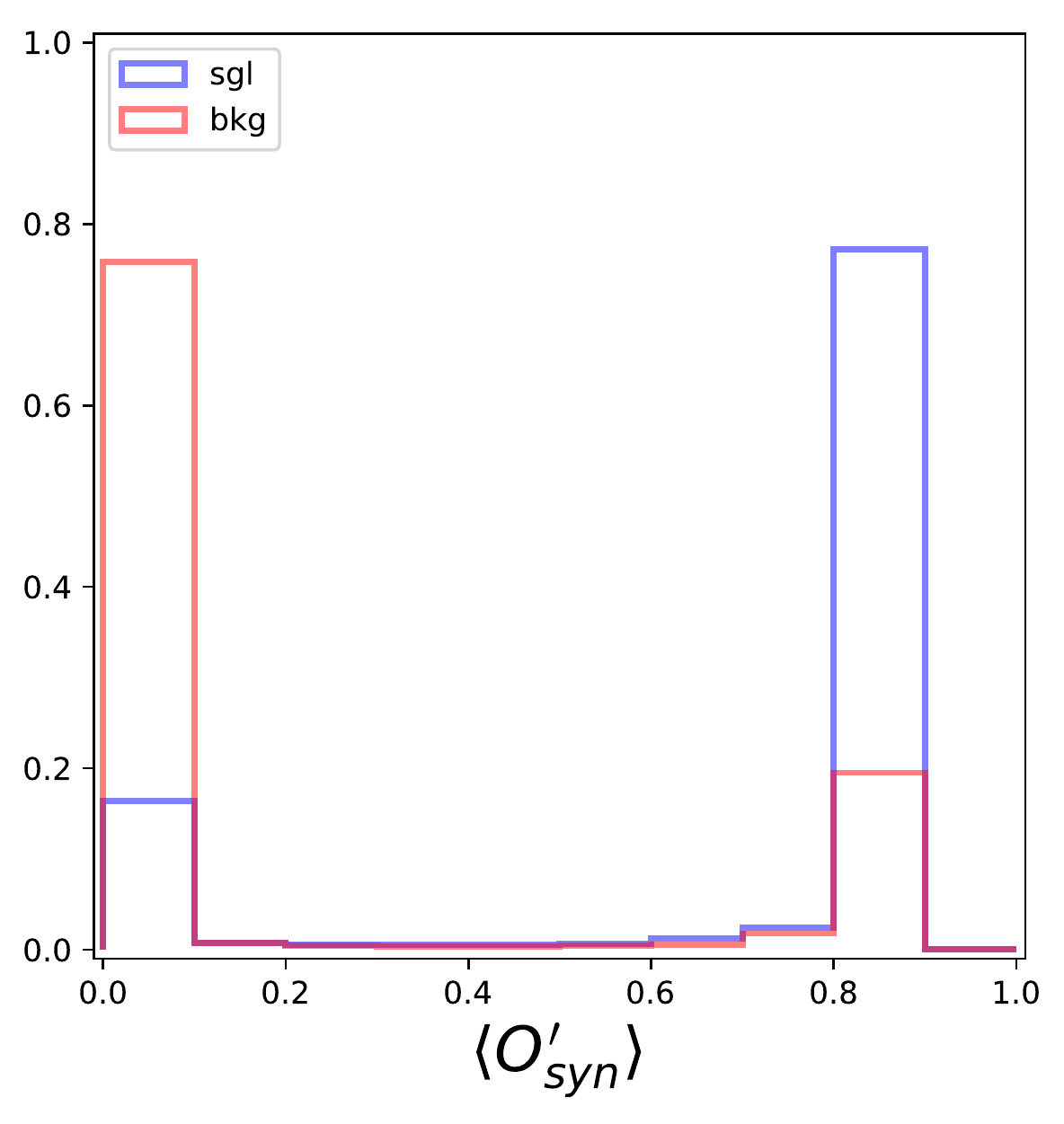}
\caption{}
\end{subfigure}
\caption{Distribution of scores according to the different $R_{\rm AE}$-based novelty evaluators: (a) $\mathcal O_{\rm iso}$, (b) $\mathcal O_{\rm clu}$, (c) $\mathcal O_{\rm syn}$ and (d) $\langle O'_{\rm syn} \rangle$, for the 2D Gaussian samples corresponding to BP 5.}
\label{fig:2ae_O}
\end{figure} 

\begin{figure}[ht]
\centering
\begin{subfigure}[b]{0.45\textwidth} 
\centering
\includegraphics[width=\textwidth]{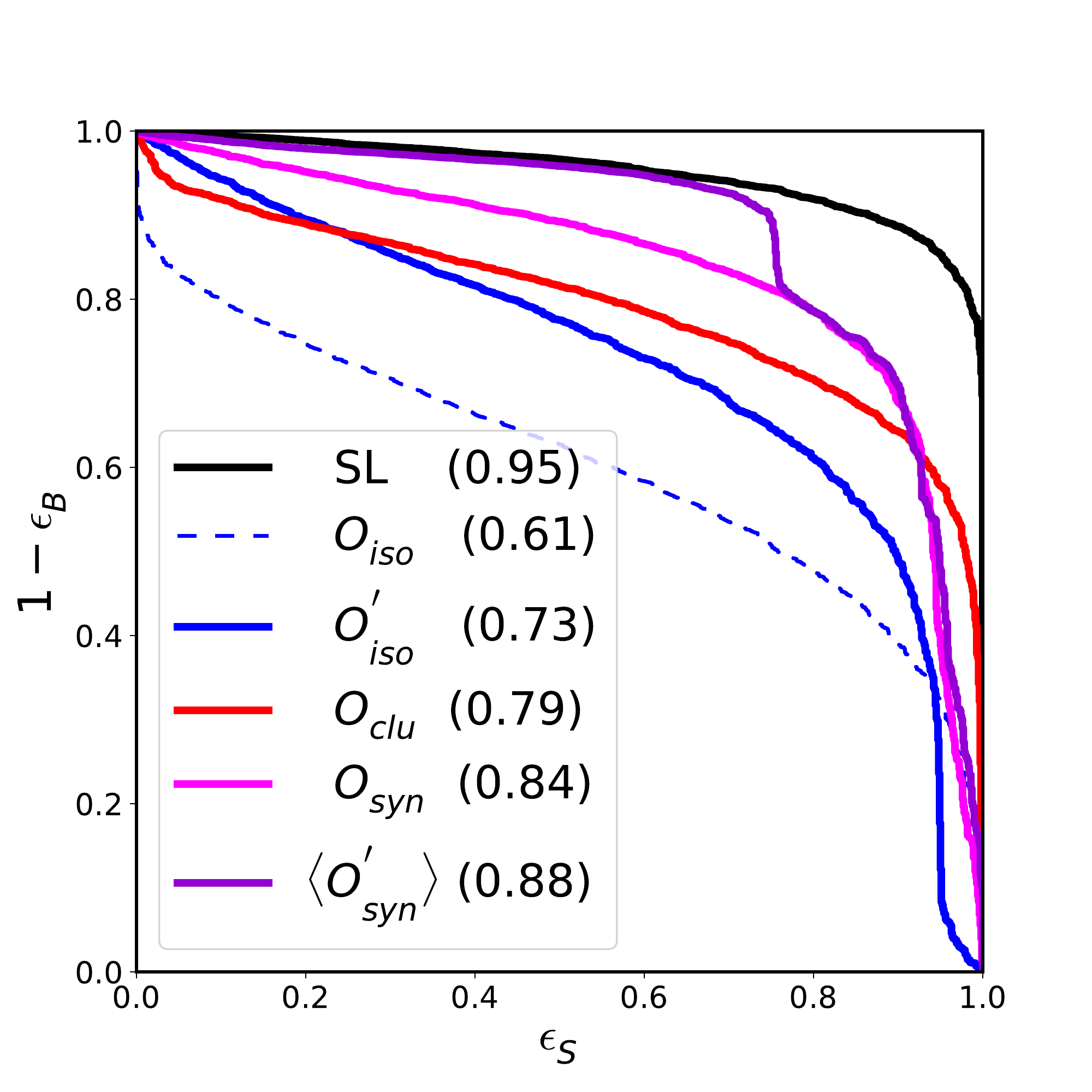}
\caption{}
\end{subfigure}
\begin{subfigure}[b]{0.45\textwidth} 
\includegraphics[width=\textwidth]{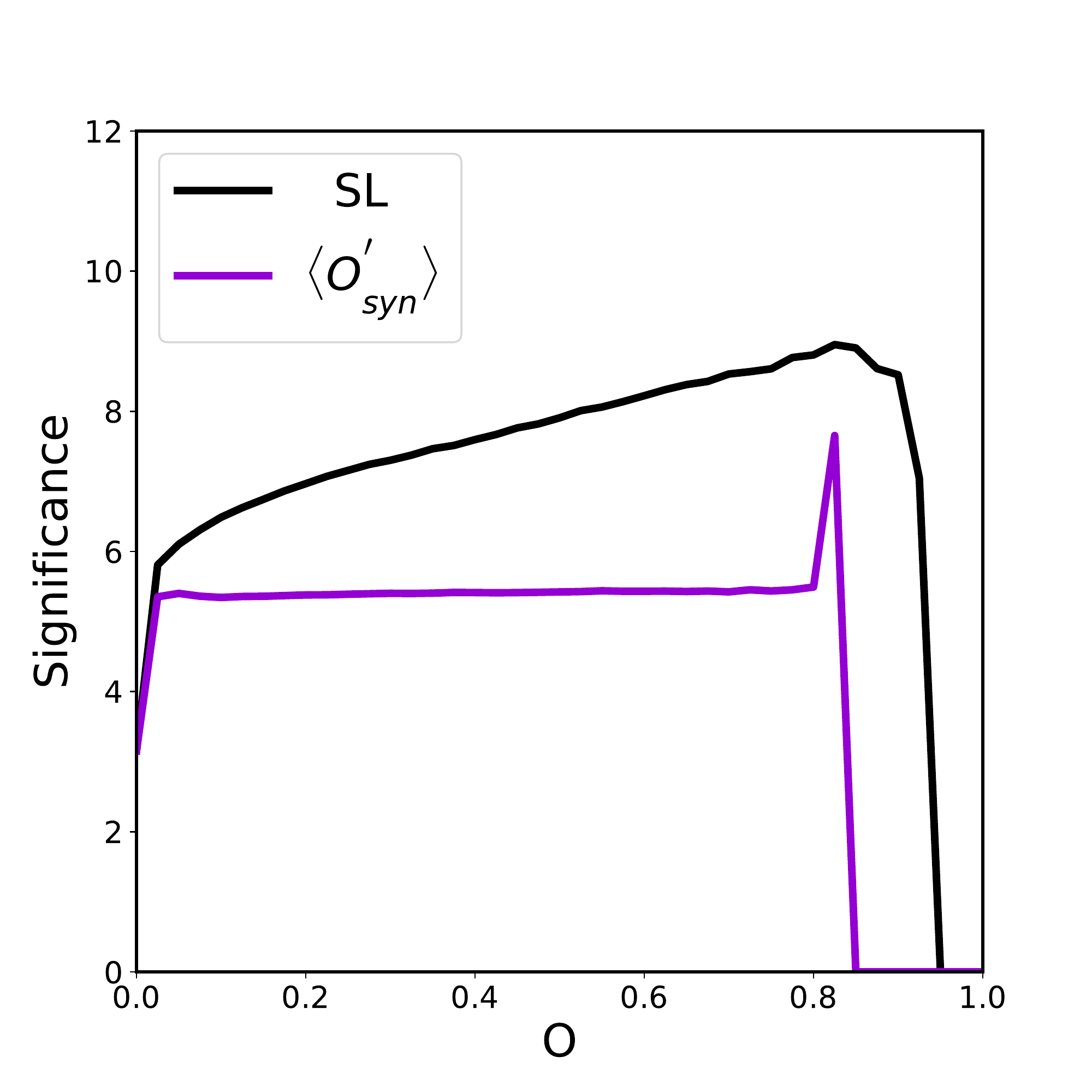}
\caption{}
\end{subfigure}
\caption{(a) ROC curves and their AUC values (given in brackets) for the set of $R_{\rm AE}$-based novelty evaluators considered and (b) statistical significance as a function of the novelty score threshold (40 bins are defined), applied to the 2D Gaussian testing samples corresponding to BP 5. Here $r_0=0.6$ is taken to define the signal-like sample $S'$ for training ${\mathcal O'_{\rm syn}}$. As a reference, the SL performance has also been reported.} 
\label{fig:2ae_case5_roc}
\end{figure} 

The novelty response to the $R_{\rm AE}$-based $\mathcal O_{\rm iso}$, $\mathcal O_{\rm clu}$, $\mathcal O_{\rm syn}$ and $\langle \mathcal{O'}_{\rm syn} \rangle$ are shown in \fig{2ae_O}. Their ROC curves and the correspondent AUC values, and the significance curve of $\langle \mathcal{O'}_{\rm syn} \rangle$ are shown in \fig{2ae_case5_roc}. Comparing with the novelty response to the $k$NN-based evaluators in the same BP (see Figs.~\ref{fig:oiso} and~\ref{fig:oclu}), it can be appreciated that these two designs share the same feature: the response of the signal events to $\mathcal O_{\rm iso}$ is moderate, while their response to $\mathcal O_{\rm clu}$ is strong. As discussed above, this is essentially determined by the definition of $\mathcal O_{\rm iso}$ and $\mathcal O_{\rm clu}$, rather than the actual distance measure used. Quantitatively, the $k$NN-based $\mathcal O_{\rm iso}$ tends to score these signal events lower, while the $R_{\rm AE}$-based one tends to score them higher. This yields a set of AUC values for the $R_{\rm AE}$-based $\mathcal O'_{\rm iso}$ and $\mathcal O_{\rm clu}$ that are closer to each other, compared to the $k$NN-based case. Finally, the $R_{\rm AE}$-based design returns a comparable $\langle \mathcal O'_{\rm syn} \rangle$ AUC value (0.88 VS. 0.90) and significance ($7.7\sigma$ VS. $8.5\sigma$) to those shown in \fig{AUC_ROC} and Fig.~\ref{fig:sigfg}. 

\subsection{Comparison with Ref.~\cite{DAgnolo:2019vbw}}

\begin{figure}[ht]
\centering
\begin{subfigure}[b]{0.45\textwidth} 
\includegraphics[width=\textwidth]{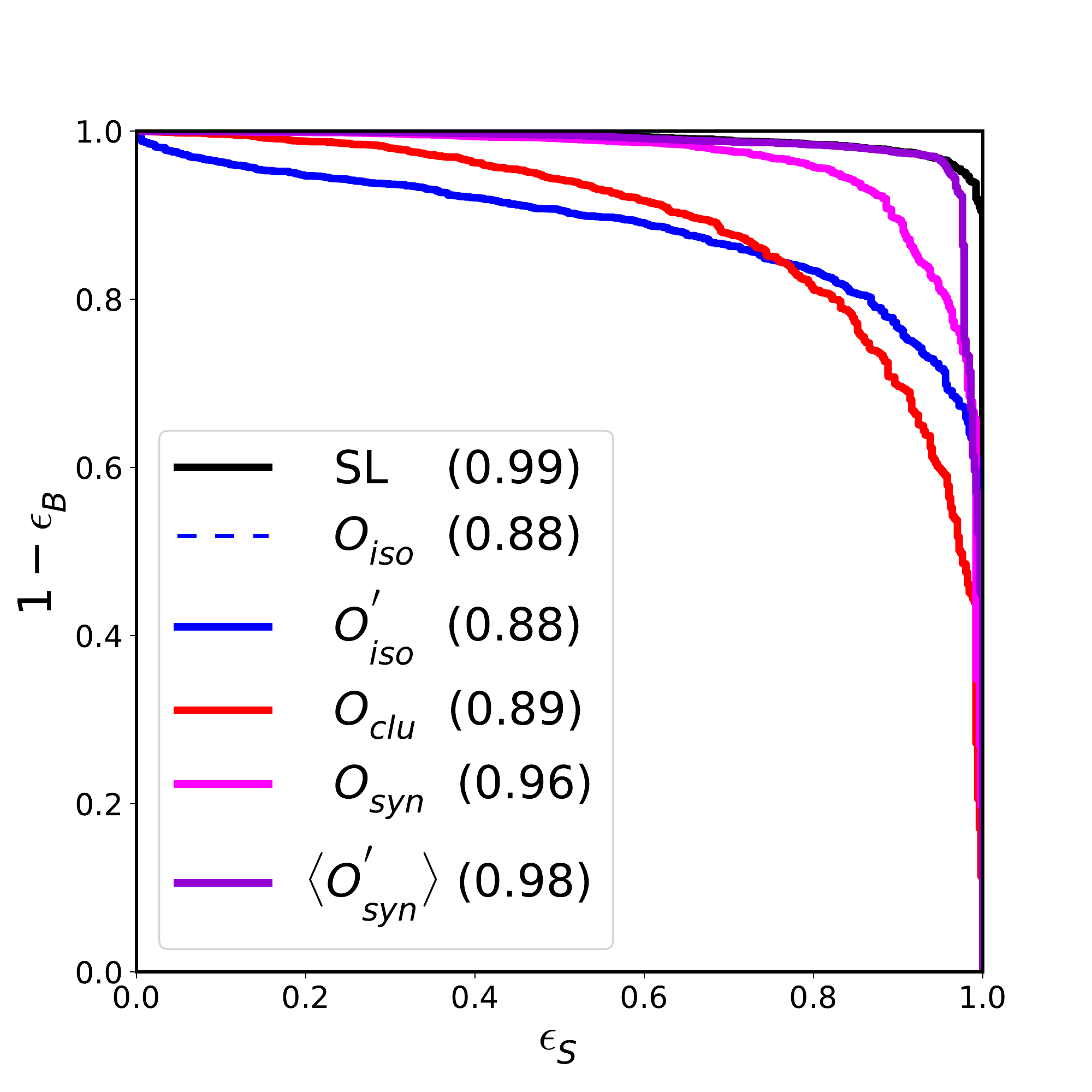}
\caption{}
\end{subfigure}
\begin{subfigure}[b]{0.45\textwidth} 
\includegraphics[width=\textwidth]{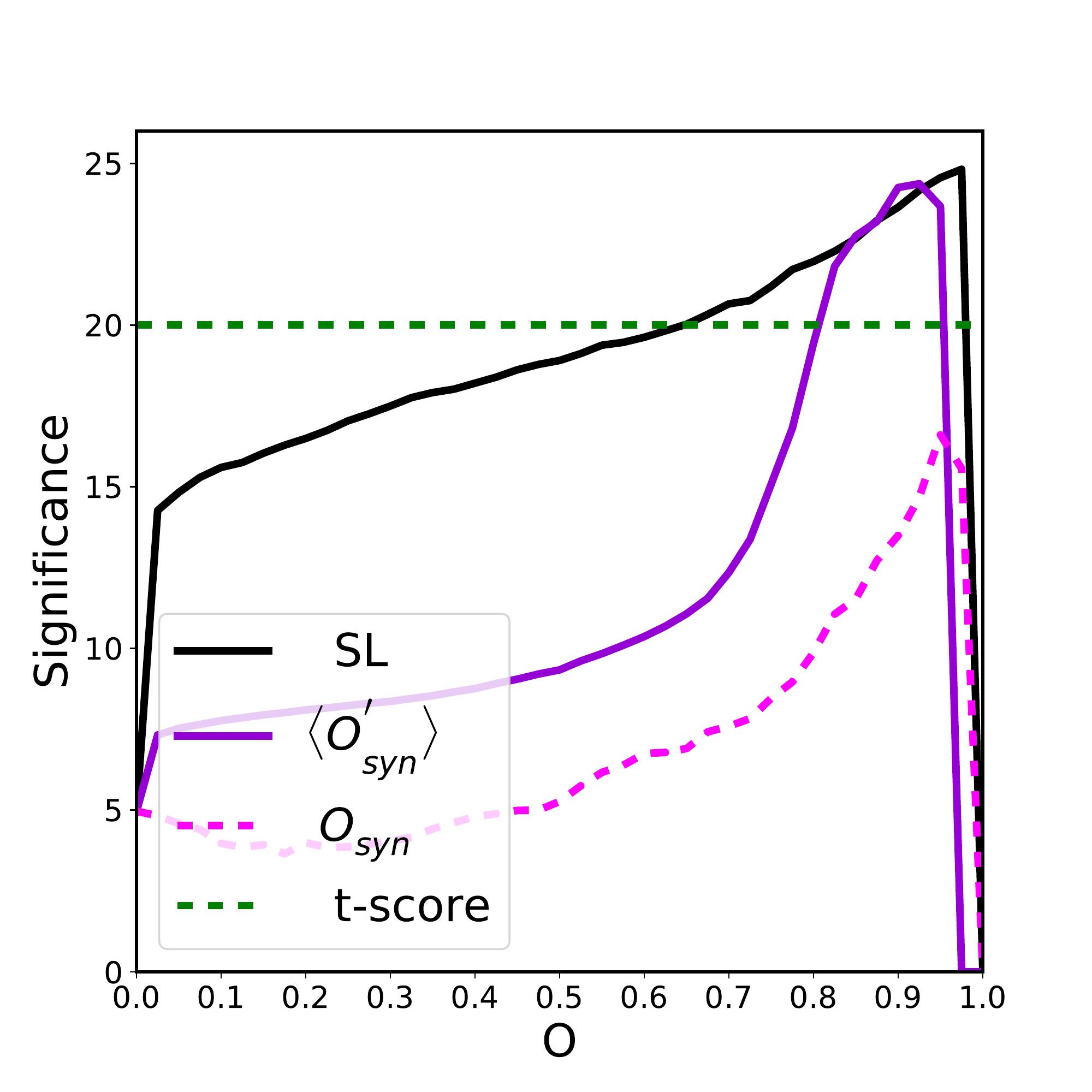}
\caption{}
\end{subfigure}
\caption{(a) ROC curves and their AUC values (given in brackets) the different novelty evaluators considered and (b) statistical significance as a function of the novelty score threshold (40 bins are defined), for the specific 2D Gaussian case studied in Refs.~\cite{Hajer:2018kqm, DAgnolo:2019vbw}. Here $r_0=0.7$ is taken to define the signal-like sample $S'$ for training ${\mathcal O'_{\rm syn}}$. As a reference, the significance obtained in \cite{DAgnolo:2019vbw} is shown in the right panel while  the SL performance is reported in both.}
\label{fig:comp}
\end{figure} 

In this subsection we consider a specific 2D Gaussian case: 10000 background events distributed according to $\mathcal{N} ((0, 0), \bf I)$) and 500 signal events according to $\mathcal{N} ((1.5, 1.5), 0.1\bf I)$. Here the reference sample contains $4\times 10^4$ background events. This case is close to BP 6 studied above, but with slightly larger (effective) $\mu$ and $\sigma$ values, and five times larger $S/B$. It was first applied in~\cite{Hajer:2018kqm} to show the capability of ${\mathcal O_{\rm iso}}$, $\mathcal{O}_{\rm clu}$ and $\mathcal{O}_{\rm syn}$ and then used in~\cite{DAgnolo:2019vbw} to demonstrate the performance of the likelihood-based algorithm~\cite{DAgnolo:2018cun}.  Here we present the ROC curves and their AUC values for the set of novelty evaluators, and the significance curve of $\langle \mathcal{O'}_{\rm syn} \rangle$ in \fig{comp}. In the original paper~\cite{Hajer:2018kqm}, a maximal significance $\sim 17\sigma$ is obtained with $\mathcal{O}_{\rm syn}$~\cite{Hajer:2018kqm}, while in Ref.~\cite{DAgnolo:2019vbw}  the significance quoted is around $20\sigma$. Using the suggested scheme in this paper, however, we find that a maximal significance $\sim 24\sigma$ can be reached by $\mathcal{O'}_{\rm syn}$. It improves the maximal significances reported in Refs.~\cite{Hajer:2018kqm,DAgnolo:2019vbw} by $\sim 7\sigma$ and $\sim 4\sigma$, respectively. Actually, the performance of $\mathcal{O'}_{\rm syn}$ in this context is very close to the maximal significance of $\sim 24.5\sigma$ obtained with SL.

\begin{figure}[ht]
\centering
\begin{subfigure}[b]{0.45\textwidth} 
\includegraphics[width=\textwidth]{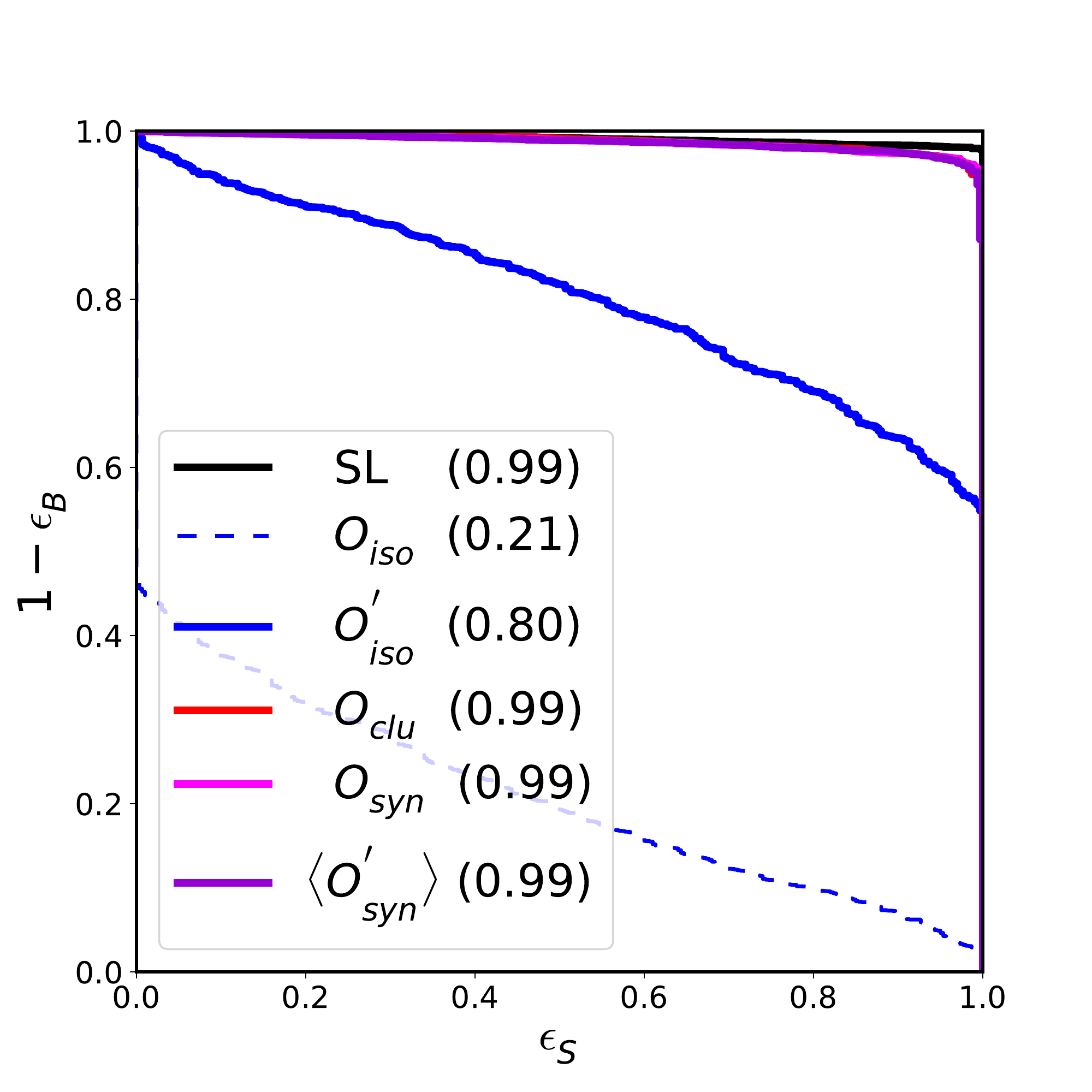}
\caption{}
\end{subfigure}
\begin{subfigure}[b]{0.45\textwidth} 
\includegraphics[width=\textwidth]{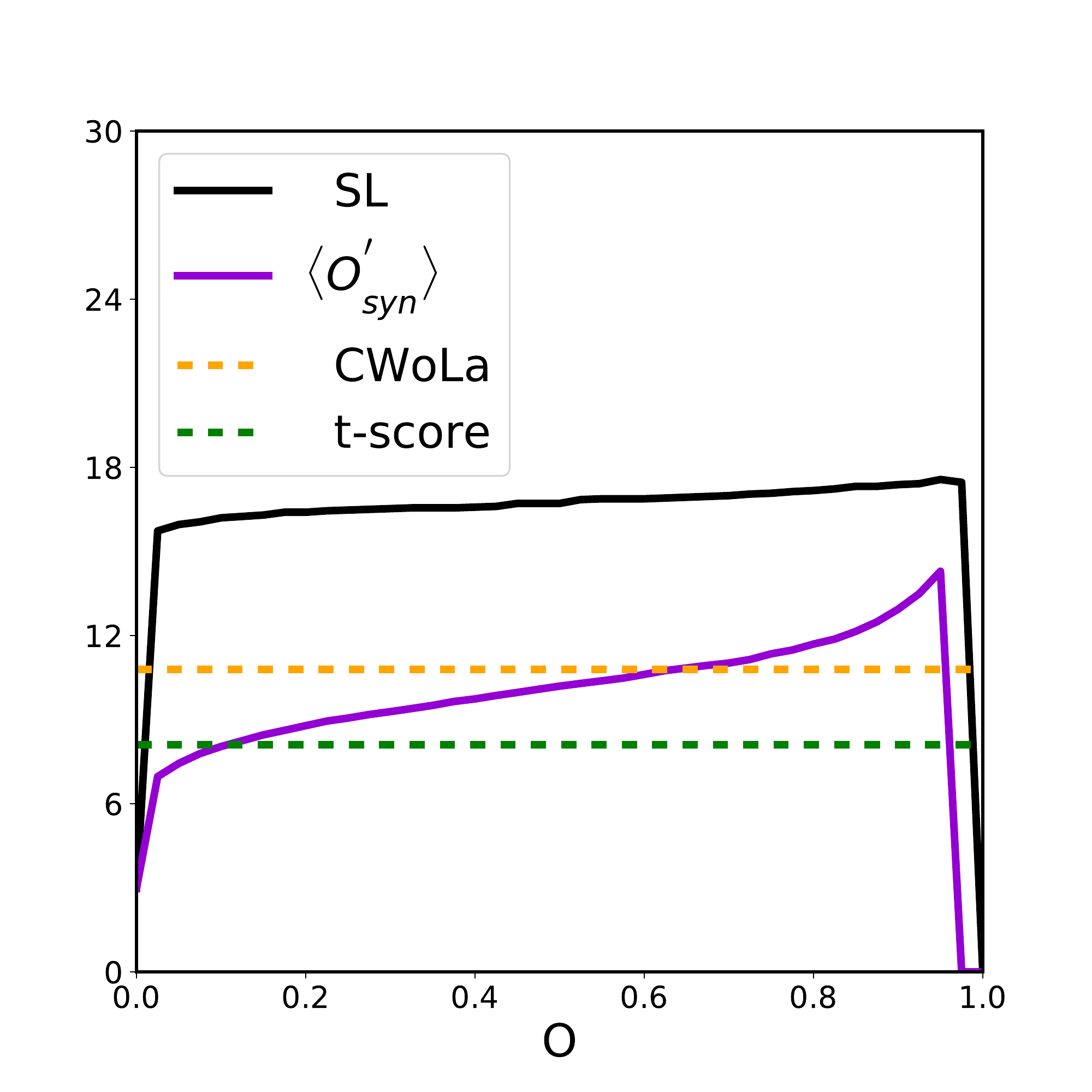}
\caption{}
\end{subfigure}
\caption{(a) ROC curves and their AUC values (given in brackets) for the different novelty evaluators considered and (b) statistical significance as a function of the novelty score threshold (40 bins are defined), for the specific case of 3D resonant bump hunting studied in Refs.~\cite{Collins:2018epr,DAgnolo:2019vbw}. Here $r_0=0.8$ is taken to define the signal-like sample $S'$ for training ${\mathcal O'_{\rm syn}}$. As a reference, the significances obtained in \cite{Collins:2018epr,DAgnolo:2019vbw} are shown in the right panel while the SL performance is reported in both.}
\label{fig:toy2}
\end{figure}

The comparison can be extended to another specific case which is on the resonant bump hunting in a 3D feature space $(m,~x,~y)$. This toy case was first considered by~\cite{Collins:2018epr} and further analyzed in~\cite{DAgnolo:2019vbw}. In this case, the signal and background events are uniformly distributed in a region defined by $(|m|<1,~|x|<0.1,~|y|<0.1)$ and $(|m|<2,~|x|<0.5,~|y|<0.5)$, respectively. The reference sample has 40000 background events and the testing sample consists of 10000 background events and 300 signal events. We present the ROC curves and their AUC values for the set of novelty evaluators, and the significance curve of $\langle \mathcal{O'}_{\rm syn} \rangle$ in \fig{toy2}. Based on $\mathcal{O'}_{\rm syn}$, we find a maximal significance $\sim 13 \sigma$ which is not far from the SL reach. As a comparison, the $t$-score method gives a median global significance of 8.1$\sigma$~\cite{DAgnolo:2019vbw} and the CWoLa hunting yields a local significance $\sim 10.8\sigma$~\cite{Collins:2018epr}. 

Finally, we would point out that these comparisons are based on some simple toy cases known to us. To have a full picture on the performance of these methods, one needs to go to the cases with more realistic and dedicated kinematics. We leave this to a future work. Instead, below we will apply the suggested scheme to the analysis of $t\bar{t}\gamma\gamma$ events in the proton-proton collider.

\section{A Collider Case Study: Novelty Detection in $t\bar{t}\gamma\gamma$ Events} 
\label{sec:S3}
In this section, we will apply the proposed novelty detection scheme to a more realistic case: the analysis of $t\bar{t}\gamma\gamma$ events produced in proton-proton collisions at $\sqrt{s}=13$~TeV, assuming an integrated luminosity of 3 ab$^{-1}$. The potential signal events to detect include: (1) SM $t \bar t h$ production with $h\to \gamma\gamma$, and (2) 
direct stop-quark pair production ($\tilde t \bar{\tilde t}$) in gravity-mediated SUSY, with both stop quarks undergoing a chain decay $\tilde t \to t {\tilde \chi}_1^0 (\to \gamma \tilde G)$, resulting in large missing transverse momentum from the undetected gravitinos ($\tilde G$). They represent two typical signatures at colliders: a resonant peak and a broad shape. As will be shown, these two types of signal events have a strong response to $\mathcal{O}_{\rm clu}$ and $\mathcal{O}_{\rm iso}$, respectively. But eventually both of them can be picked up by $\mathcal{O}'_{\rm syn}$ with a higher efficiency.  So this analysis provides a nice context to test the suggested analysis scheme. As for the backgrounds, they arise primarily from $t\bar{t}\gamma\gamma$, $t\bar{t}\gamma$+jets (with one jet misidentified as a photon), $t\bar{t}$+jets (with two jets misidentified as a photon), and continuum $\gamma\gamma$+jets production.  

\subsection{Event Generation}
Monte Carlo (MC) event samples are generated with {\sc MadGraph5}\_{\sc aMC@NLO} 2.6.6~\cite{Alwall:2014hca} and showered with {\sc Pythia} 8~\cite{Sjostrand:2007gs}. We use {\sc Delphes} 3.4.1~\cite{deFavereau:2013fsa} to simulate the ATLAS detector response, obtaining similar SM $t \bar t h$ yield as in Ref.~\cite{ATLAS:2019aqa}. We apply an event preselection following Ref.~\cite{Aaboud:2018xdt}, by requiring at least two photons ($p_T > \unit[25]{GeV}$, $|\eta| < 2.37$ but with an exclusion of $1.37 < |\eta| < 1.52$), exactly one isolated electron or muon ($p_T > \unit[10]{GeV}$, $|\eta| < 2.7$ but with an exclusion of $1.37 < |\eta| < 1.52$ for electrons), at least two central jets ($p_T > \unit[25]{GeV}$ and $|\eta| < 2.4$) and at least one $b$-tagged jet with a $b$-tagging efficiency of $70\%$.

The background samples are simulated at leading order, following Ref.~\cite{Aaboud:2018xdt}. We use the MLM algorithm to perform the matrix-element to parton-shower matching for the $t\bar{t}\gamma$ process up to one additional jet, and for the $t\bar t$ process up to two additional jets. Re-weighting, with the probability provided in Ref.~\cite{Homiller:2018dgu}, is applied to compensate for the low-generation efficiency caused by the tiny faking rate of photons by jets. The yields after preselection for $t\bar{t}\gamma\gamma$, $t\bar{t}\gamma$ and $t\bar t$ are eventually close to $78.8:18.6:2.6$~\cite{Aaboud:2018xdt}. The continumm $\gamma\gamma$ process contributes about half of the total backgrounds~\cite{ATLAS:2019aqa}. Here it is simulated in a five-flavour scheme, with MLM matching up to two additional jets. One of these jets is randomly assigned in each event to represent the jet misidentified as a lepton, as required by the preselection. 

The $t \overline t h$ signal sample is generated at NLO, forcing the $h \to \gamma\gamma$ decay, and normalized to a cross section of $\unit[0.51]{pb}$ times the Higgs boson branching ratio to
diphotons of 0.227\%~\cite{Aaboud:2018xdt}. The $\tilde t \bar{\tilde t}$ sample is generated at leading order, with the stop quark ($\tilde t$) decaying into a top quark and a bino-like neutralino ($\tilde{\chi}_1^0$) and this neutralino further decaying into a photon and a gravitino ($\tilde{G}$). The mass parameters for stop quark, neutralino and gravitino are set as $m_{\tilde{t}} = \unit[1]{TeV}$, $m_{\tilde{\chi}_1^0} = \unit[0.2]{TeV}$ and $m_{\tilde{G}} \sim \unit[0]{GeV}$. The $\tilde t \bar{\tilde t}$ sample is normalized to a cross section of $\unit[6.83]{fb}$~\cite{Borschensky:2014cia}, times an assumed branching fraction of 10\% for decay into $t\bar{t}\gamma\gamma + 2\tilde{G}$. 

The expected event yields after preselection are summarized in Table~\ref{tab:yield}. In addition, we generate $1.3\times 10^5$ background events satisfying the preselection criteria as the reference sample.

\begin{table} [H]
 \begin{center}
  \begin{tabular}{c|c|c|c}
  \hline
 & Process &  Matching &  Event yields \\ \hline
\multirow{4}{*}{Backgrounds} &$t\bar t\gamma\gamma$ &No& 765 \\ \cline{2-4}
& $t\bar t \gamma $ & Up to one jet & 370 \\ \cline{2-4}
& $t\bar t $& Up to two jets & 83 \\ \cline{2-4}
&Continuum $\gamma\gamma$& Up to two jets  & 1216 \\ \hline
$t\bar t h$ &$t\bar th(\gamma\gamma)$ & No & 167 \\ \hline
SUSY &$\tilde{t}\tilde{\bar{t}}\to t\bar{t}\gamma\gamma + 2\tilde{G}$ & No & 226 \\ \hline
\end{tabular}
\end{center}
\caption{Expected events yields after preselection, assuming 3 ab$^{-1}$ at $\sqrt{s}=13$ TeV. } 
\label{tab:yield}
\end{table}

\subsection{Novelty Analysis}
Although even after preselection multiple kinematic features could be exploited to discriminate the signals from the background, in this analysis we will only consider the di-photon kinematics for simplicity. We sort the two photons by energy and take their four momentum \{$P_T, \eta, \phi, E$\} to define the eight-dimensional feature space. We then standarize each of these features as
\begin{eqnarray}
x \to \frac{x - \mu_x}{\sigma_x} \ ,
\label{eq:standard}
\end{eqnarray}
where $\mu_x$ is the mean of the reference sample and $\sigma_x$ is its variance. To reduce the potential sparse errors caused by the low rate of rare events, we take a dimensionality reduction by encoding the data in the eight-dimenional feature space into the AE latent space~\cite{Hajer:2018kqm}. The AE is built with eleven layers with 8, 12, 8, 8, 6, 2, 6, 8, 8 , 12, 8 neurons respectively. The resulting latent space is two-dimensional~\footnote{The hidden-layer architecture could be further optimized. However, given the relative simple signatures for the signals considered here, encoding the di-photon kinematics into a latent space with more than two dimensions would not help much as we have tested.}. We choose Tanh to be the activation function except for the last layer, where a linear activation function is applied to ensure the matching of the output and input data ranges. The model is then trained with a batch size of 250, a learning rate of 0.94 and a decay rate of 0.88, with ADADELTA~\cite{zeiler2012adadelta} being used as an optimizer.

The directions of the latent space rely on the definition of the loss function. As a physical requirement, we incorporate Lorentz invariants such as the single-photon and di-photon invariant masses into the vanilla loss function ($i.e.$ $L = \left | x - x' \right |^2$), in the framework of relational AE~\cite{Meng_2017}. Then we have 
\begin{eqnarray}
L' &=& L  + c \times \left ( \frac{m_{\gamma_1}^{\prime 2}}{\sigma_{E_1}^2} + \frac{m_{\gamma_2}^{\prime 2}}{\sigma_{E_2}^2} \right ) + c \times \frac{(m_{\gamma\gamma} - m_{\gamma\gamma}^{\prime})^2}{\sigma_{m_{\gamma\gamma}}^2}  \ .
\label{eq:lossfunc}
\end{eqnarray}
Here the input and output quantities are distinguished with ``prime''. The two photons are sorted by energy, with $m_{\gamma_i}$ and $m_{\gamma\gamma}$ being the single photon and di-photon invariant masses, and $\sigma_{E_i}$ and $\sigma_{m_{\gamma\gamma}}$ being the single-photon energy and di-photon $m_{\gamma\gamma}$ variances (in the background sample). To ensure each term in Eq.~(\ref{eq:lossfunc}) contributes comparably to the total loss, the coefficient $c$ is selected to be $10$. With such a construction, this AE not only learns to reconstruct the low-level observables like single-object four-momentum, but also keeps the high-level features, such as $m_{\gamma\gamma}$, which maintain the correlation among the objects in each event. 

\begin{figure}[ht]
\centering
\begin{subfigure}[b]{0.42\textwidth} 
\includegraphics[width=\textwidth]{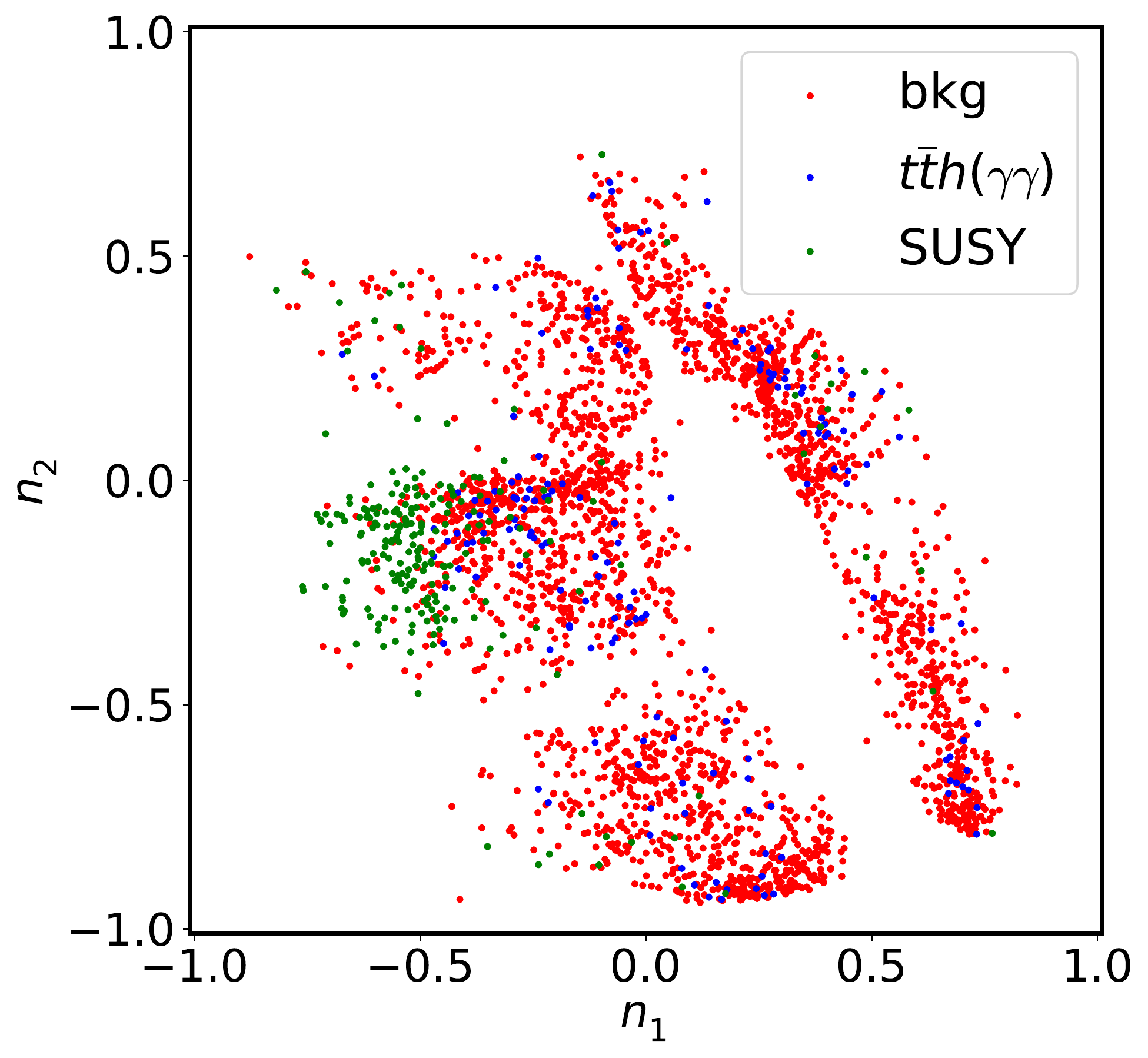}
\caption{}
\end{subfigure}
\begin{subfigure}[b]{0.42\textwidth} 
\includegraphics[width=\textwidth]{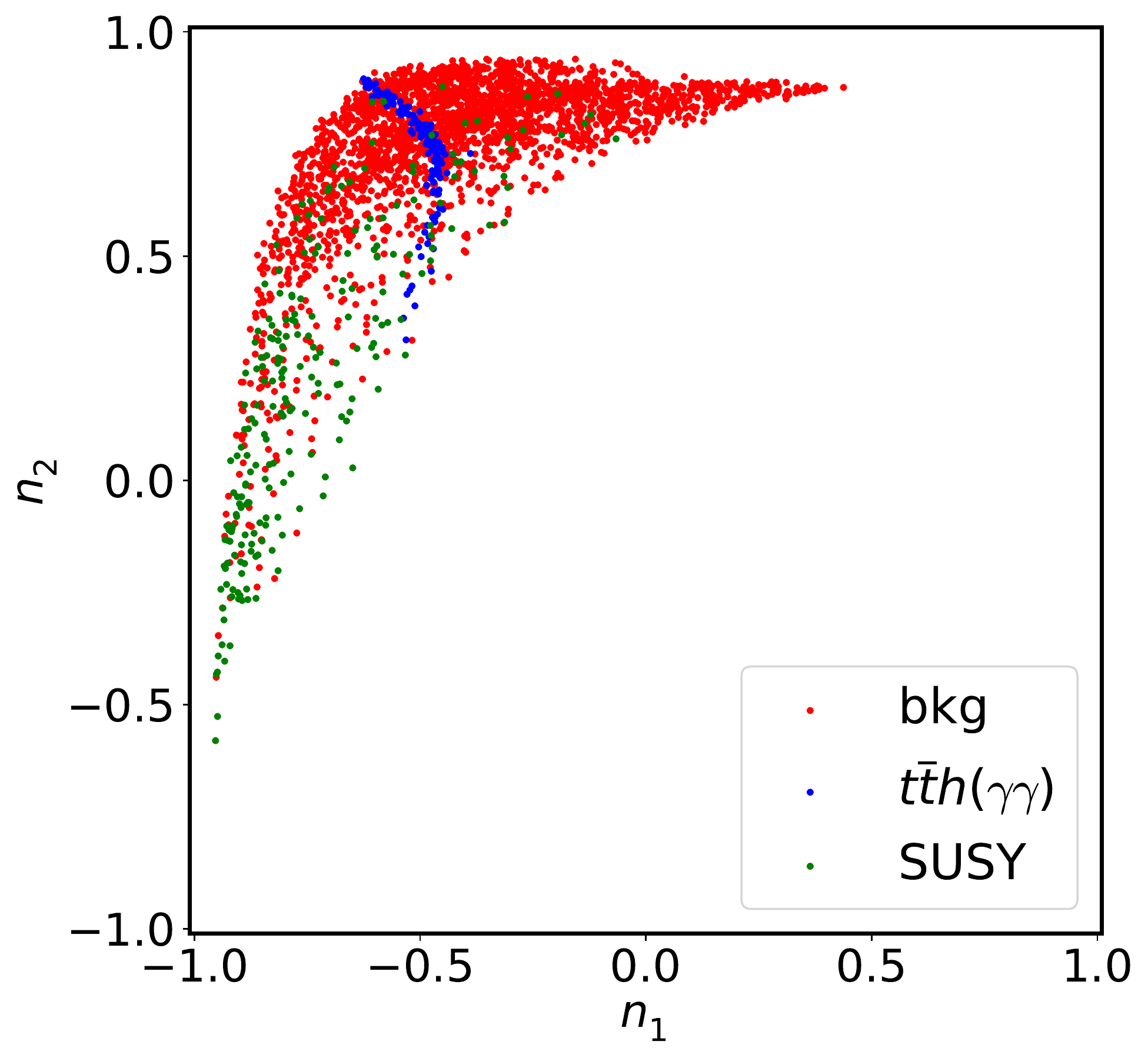}
\caption{}
\end{subfigure}
\caption{Distributions for $t\bar{t}h(\gamma\gamma)$ signal (blue points), $\tilde t \bar{\tilde t}$ signal (green points), and background events (red points) in the 2D latent space, obtained by training using either (a) the vanilla loss function $L$, or (b) the modified loss function $L'$.}
\label{fig:latent}
\end{figure}

The 2D latent spaces that are obtained by training with two different loss functions, $L$ and $L'$, are displayed in \fig{latent}. The broadly scattering of the points in the $n_1-n_2$ planes indicates that the correlation between the two dimensions of the latent space is not strong. As can be appreciated, with the Lorentz invariants being incorporated in the $L'$ loss function, the AE tends to project the signal events with different topologies into different regions in the latent space. The $t\bar t h$ signal events are highly clustered in the bulk of the background distribution while the $\tilde t \bar{\tilde t}$ signal events are broadly distributed at its tail. The internal correlations among the particles in each event, secured by Lorentz invariance, thus provide a useful guide to the AE for separating the signals with different topologies~\footnote{More generally, it will be highly valuable to embed intrinsic symmetries in particle physics (e.g., Lorentz symmetry) or implement the relevant constrains in the DNN model, to improve the effectiveness and efficiency of detecting new physics as novelty in the realistic data analyses. The exploration of this topic however is beyond the scope of this study.}. In contrast, such a symmetry-based guide is missing with the regular loss function $L$, which yields a broad distribution of the $t\bar t h$ signal events in the latent space (although the $\tilde t \bar{\tilde t}$ signal events are still loosely clustered at the tail of the background distribution). Given the similarity (in terms of both broadness and overlapping with the backgrounds) of this distribution to that of the 2D Gaussian sample in BP8, we expect the detection sensitivity of $t\bar t h$ to be low in this case. Below we will perform our analysis in the latent space defined by $L'$.

\begin{figure}[ht]
\centering
\begin{subfigure}[b]{0.42\textwidth} 
\includegraphics[width=\textwidth]{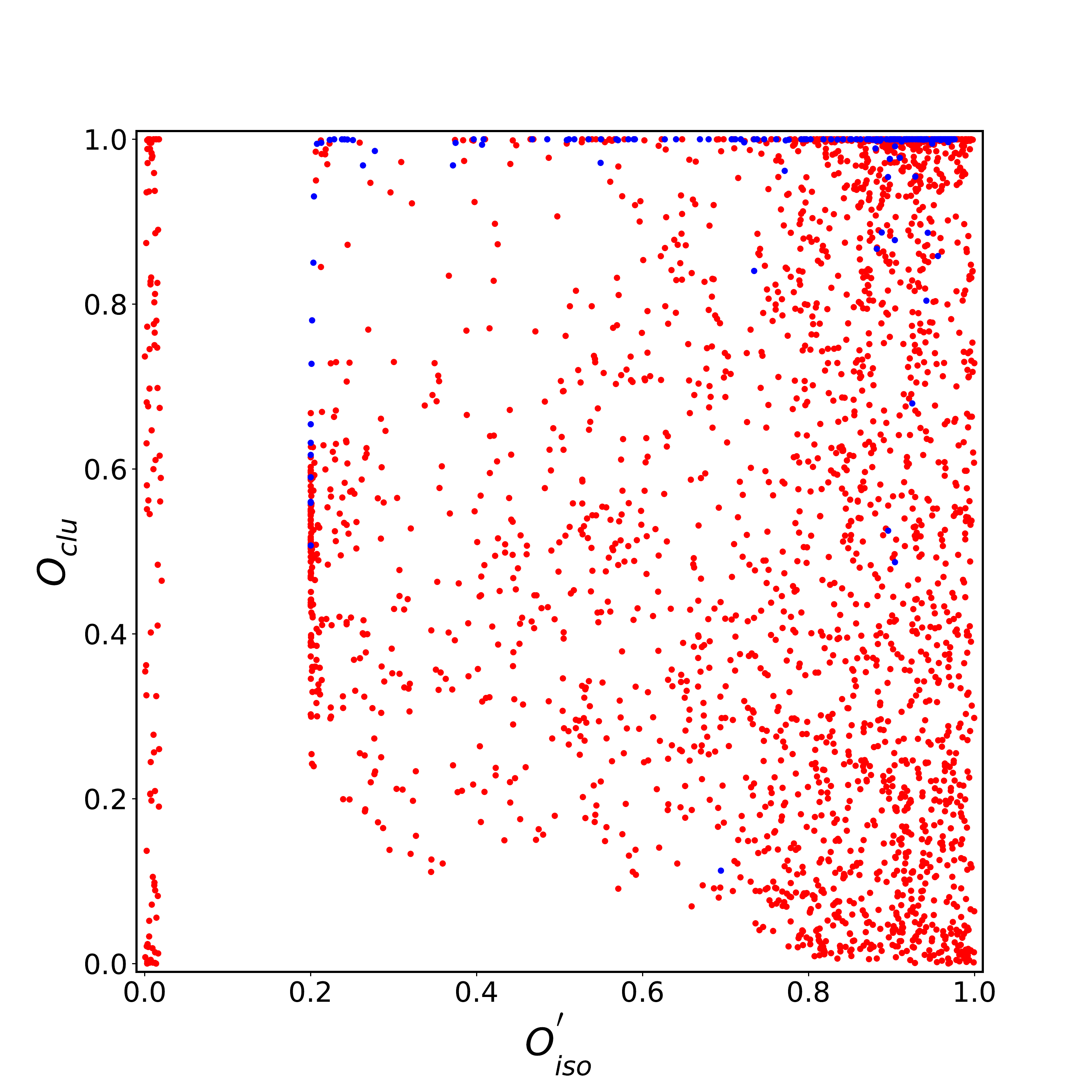}
\caption{$t\bar t h$}
\label{fig:tthscore}
\end{subfigure}
\begin{subfigure}[b]{0.42\textwidth} 
\includegraphics[width=\textwidth]{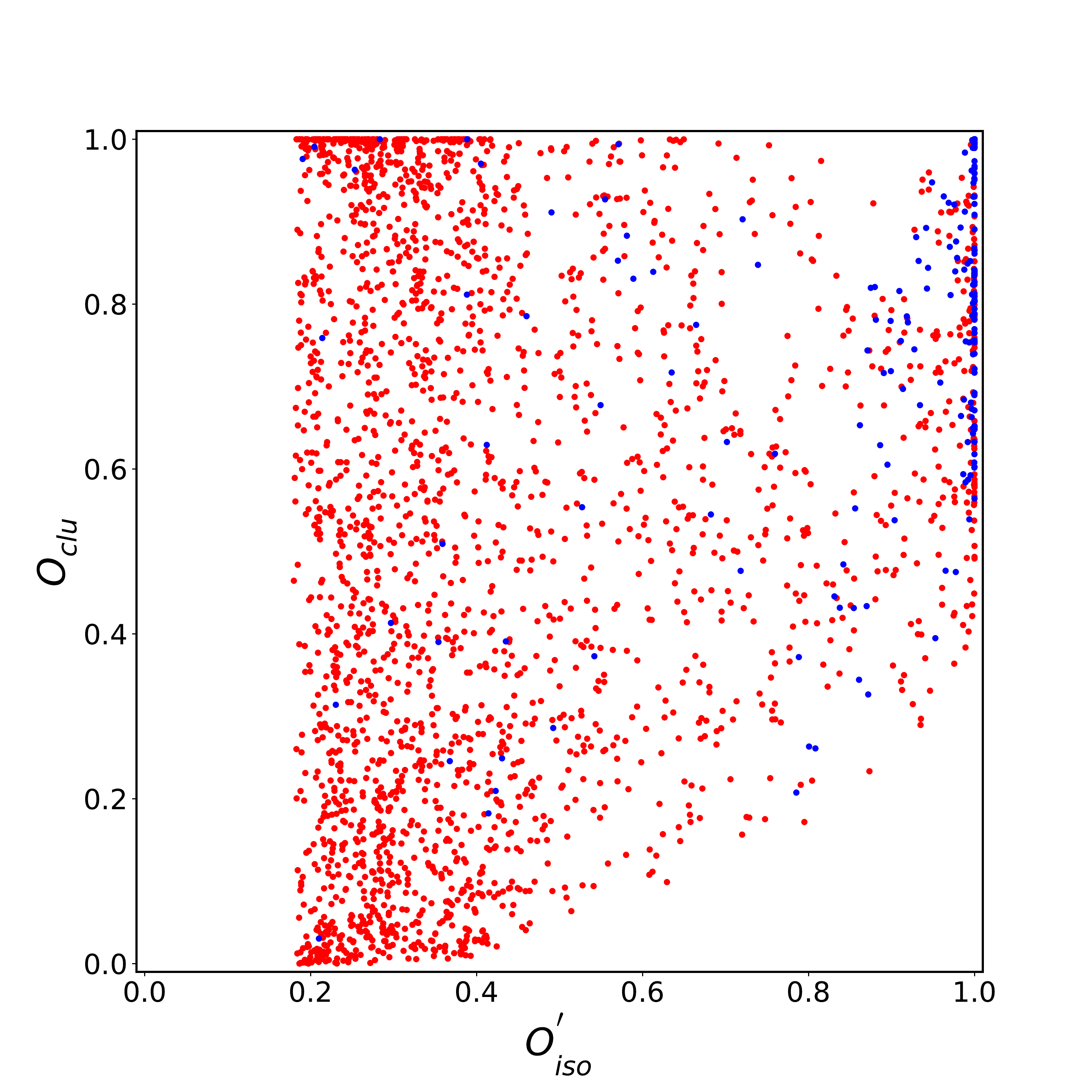}
\caption{$\tilde t \bar{\tilde t}$}
\label{fig:susyscore}
\end{subfigure}
\caption{Distribution of novelty scores for signal (blue points) and background (red points) in the $\mathcal{O}'_{\rm iso}- \mathcal{O}_{\rm clu}$ plane. The signal events considered  correspond to (a) $t\bar t h$ and (b) $\tilde t \bar{\tilde t}$. In the latter case, the bin resorting for $\mathcal{O}_{\rm iso}$ is trivial and hence we have $\mathcal{O}'_{\rm iso}=\mathcal{O}_{\rm iso}$.}
\label{fig:spacedis}
\end{figure}   

\begin{figure}[ht]
\centering
\begin{subfigure}[b]{0.42\textwidth} 
\includegraphics[width= \textwidth]{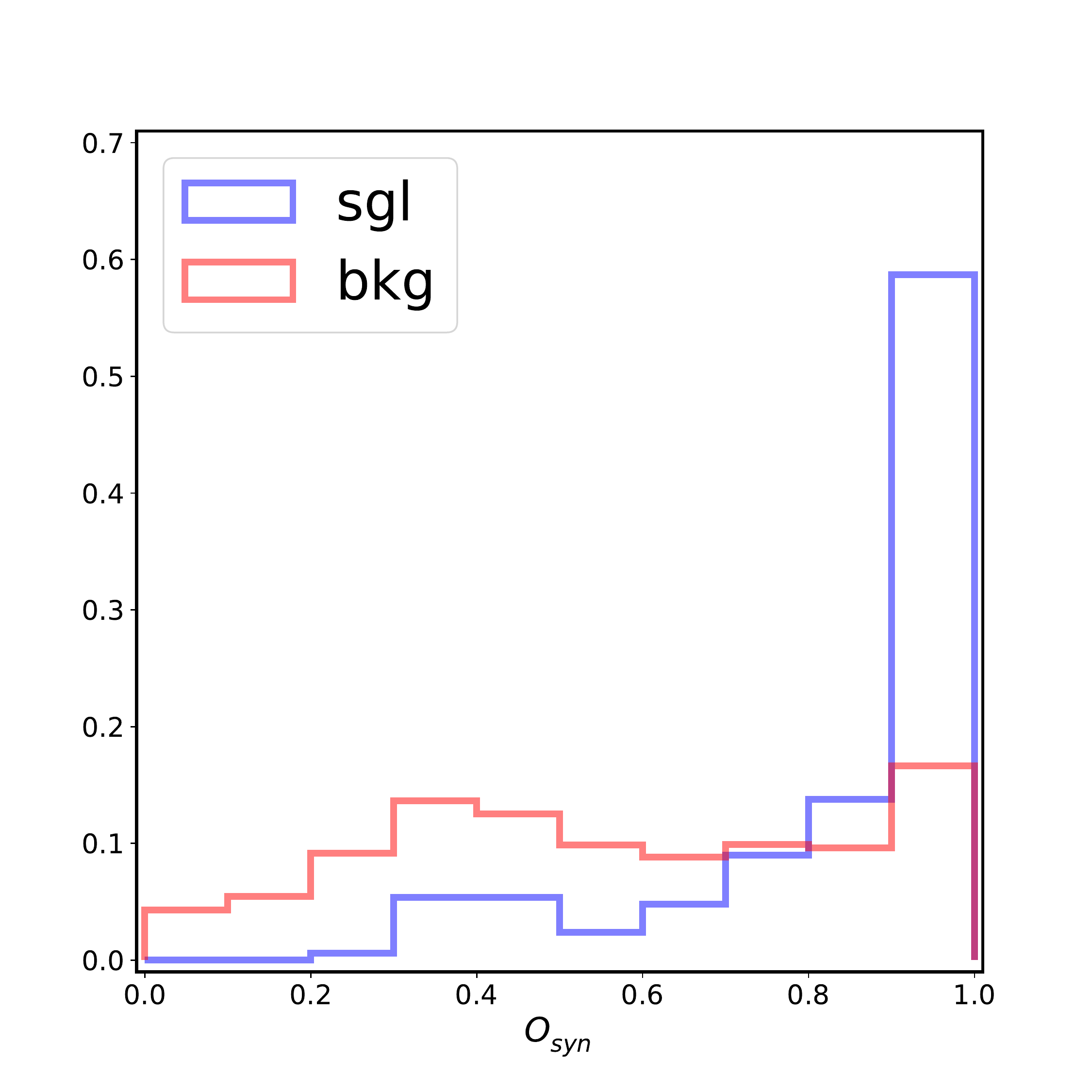}
\caption{$t\bar t h$}
\label{}
\end{subfigure}
\begin{subfigure}[b]{0.42\textwidth} 
\includegraphics[width= \textwidth]{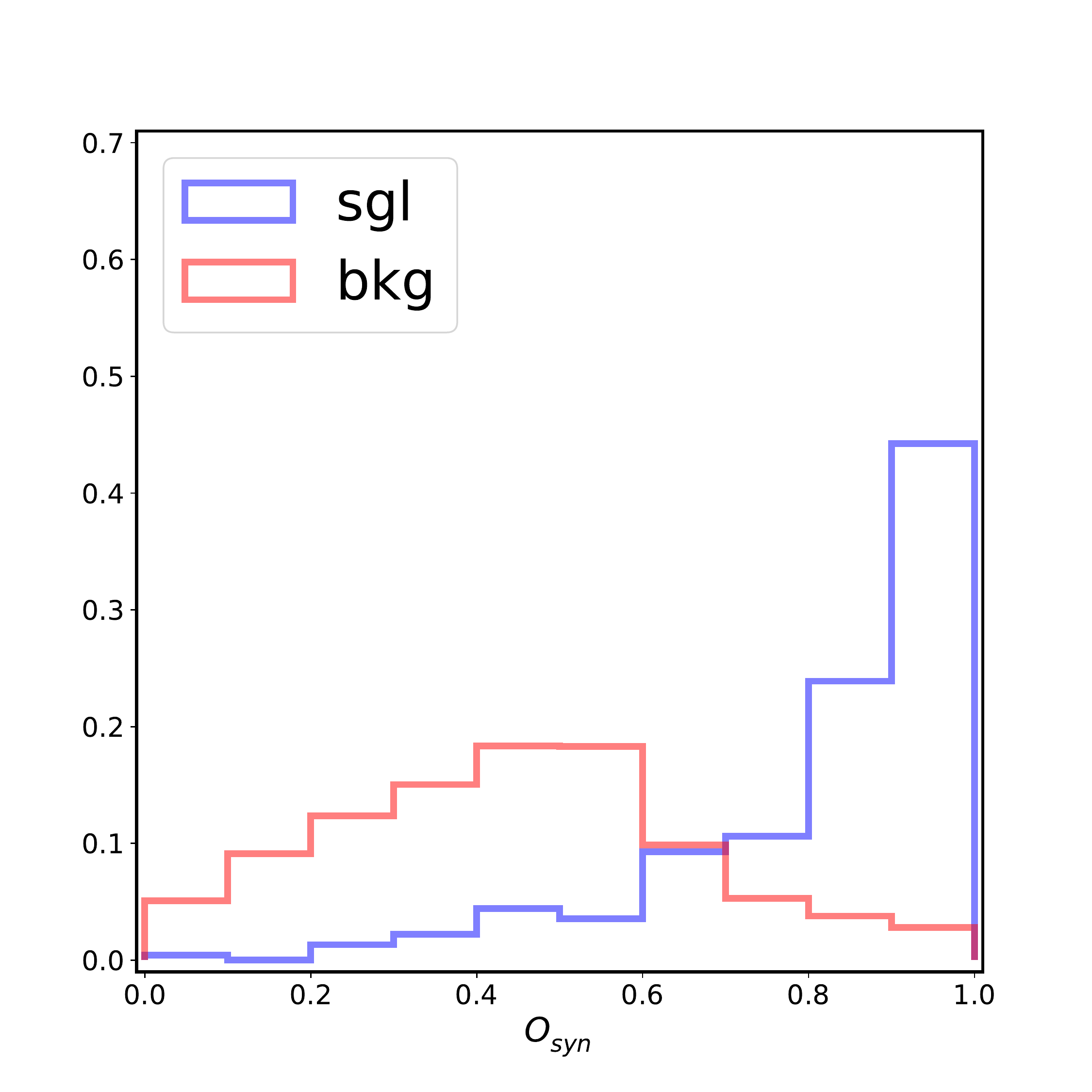}
\caption{$\tilde t \bar{\tilde t}$}
\label{}
\end{subfigure}
\caption{Novelty score according to $O_{\rm syn}$ for the signal and background events in the $t\bar t \gamma\gamma$ analysis. The signal events considered correspond to (a) $t\bar t h$ and (b) $\tilde t \bar{\tilde t}$.}
\label{fig:physsyn}
\end{figure}   

To evaluate the novelty response of the testing data, we take $k = 1000$ for $\mathcal{O}_{\rm iso}$ and $k =1000$ ($k = 19$) for $\mathcal{O}_{\rm clu}$. In the latter case, $k =1000$ is applied to calculate $d_{\rm train}$ while the rescaled $k = 19$ is applied to calculate $d_{\rm test}$. The distributions of signal and background events in the $\mathcal{O}'_{\rm iso}-\mathcal{O}_{\rm clu}$ plane are shown in Fig.~\ref{fig:spacedis}. As expected, the $t\bar t h$ and $\tilde t \bar{\tilde t}$ signal events tend to score high in $\mathcal{O}_{\rm clu}$ and $\mathcal{O}'_{\rm iso}$ (or $\mathcal{O}_{\rm iso}$), respectively, while their novelty responses to $\mathcal{O}_{\rm iso}$ and $\mathcal{O}_{\rm clu}$ are relatively weak. 
In spite of this, the synergy-based strategy ensures many of these signal events to be classified into the new signal-like bin $S'$. Indeed, in both cases the signal events obtain a relatively high $\mathcal{O}_{\rm syn}$ score, as shown in~\fig{physsyn}, and hence good discrimination against the backgrounds.

\begin{figure}[ht]
\centering
\begin{subfigure}[b]{0.42\textwidth} 
\includegraphics[width=\textwidth]{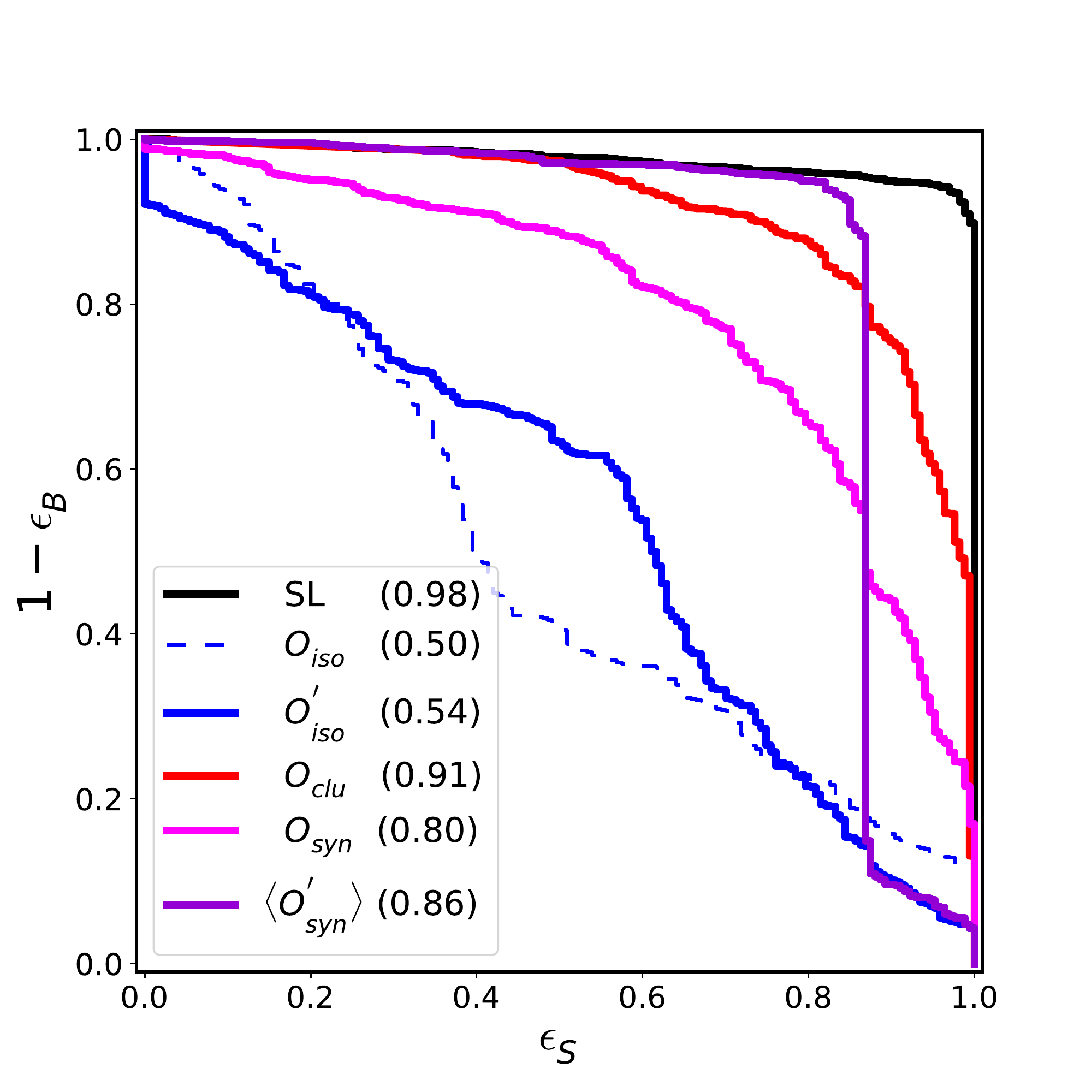}
\caption{$t\bar t h$}
\label{}
\end{subfigure}
\begin{subfigure}[b]{0.42\textwidth} 
\includegraphics[width= \textwidth]{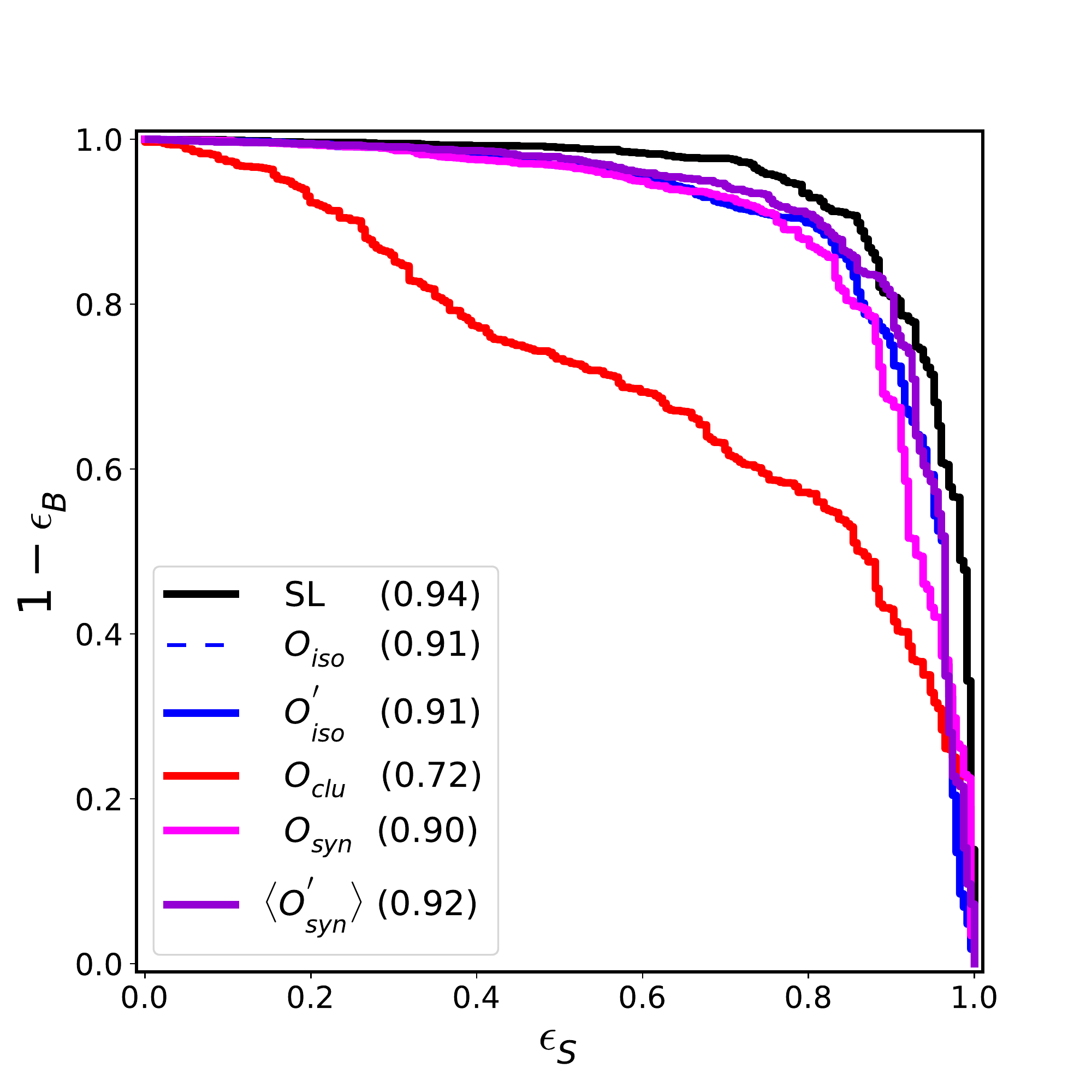}
\caption{$\tilde t \bar{\tilde t}$}
\label{}
\end{subfigure}
\caption{ROC curves and their AUC values (given in brackets) for the set of novelty evaluators considered, applied for the signal and background events in the $t\bar t \gamma\gamma$ analysis. The signal events considered correspond to (a) $t\bar t h$ and (b) $\tilde t \bar{\tilde t}$. Here $r_0=0.6$ is taken to define the signal-like sample $S'$ for training ${\mathcal O'_{\rm syn}}$. As a reference, the SL performance has also been reported.}
\label{fig:physauc}
\end{figure}  

\begin{figure}[ht]
\centering
\begin{subfigure}[b]{0.42\textwidth} 
\includegraphics[width=\textwidth]{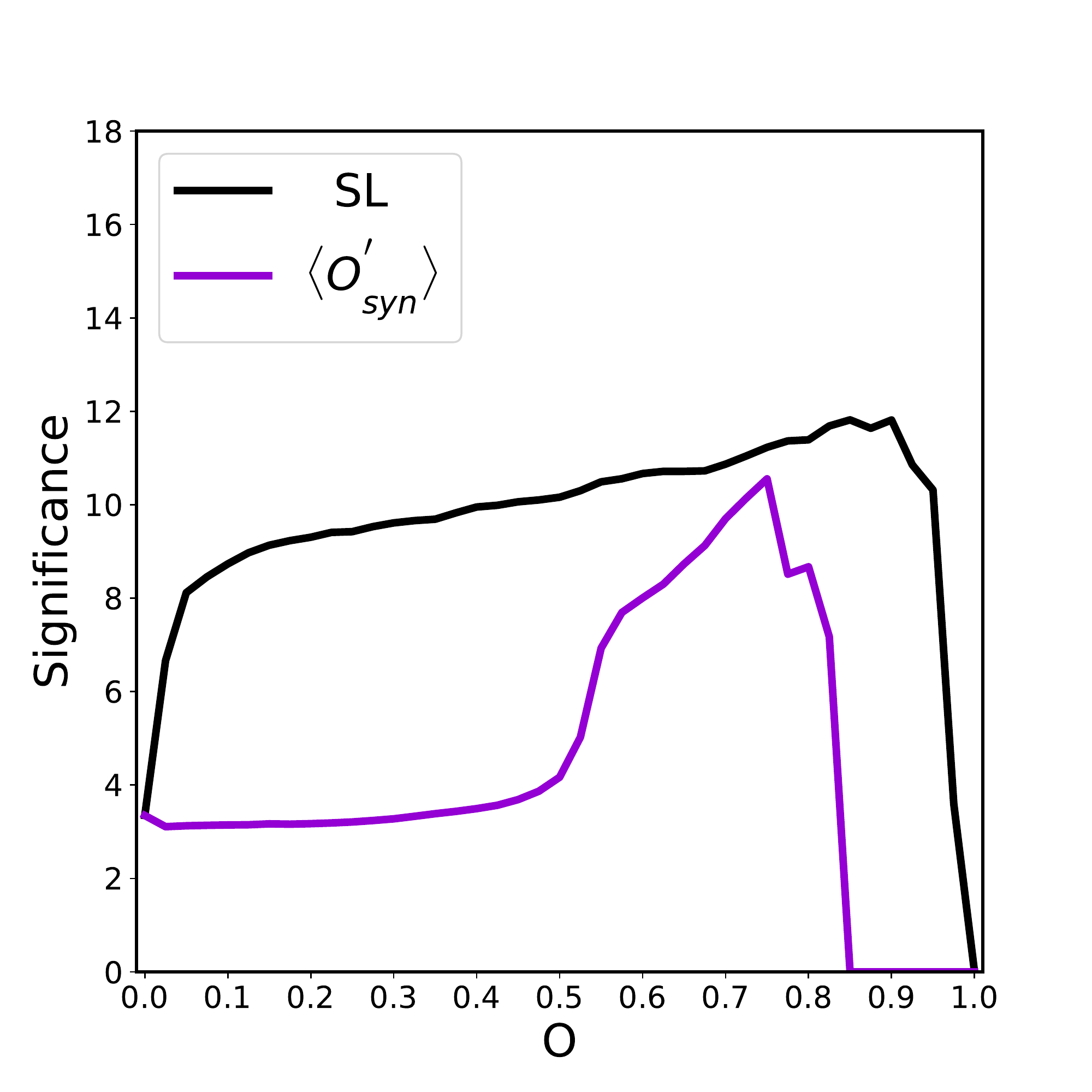}
\caption{$t\bar t h$}
\label{}
\end{subfigure}
\begin{subfigure}[b]{0.42\textwidth} 
\includegraphics[width=\textwidth]{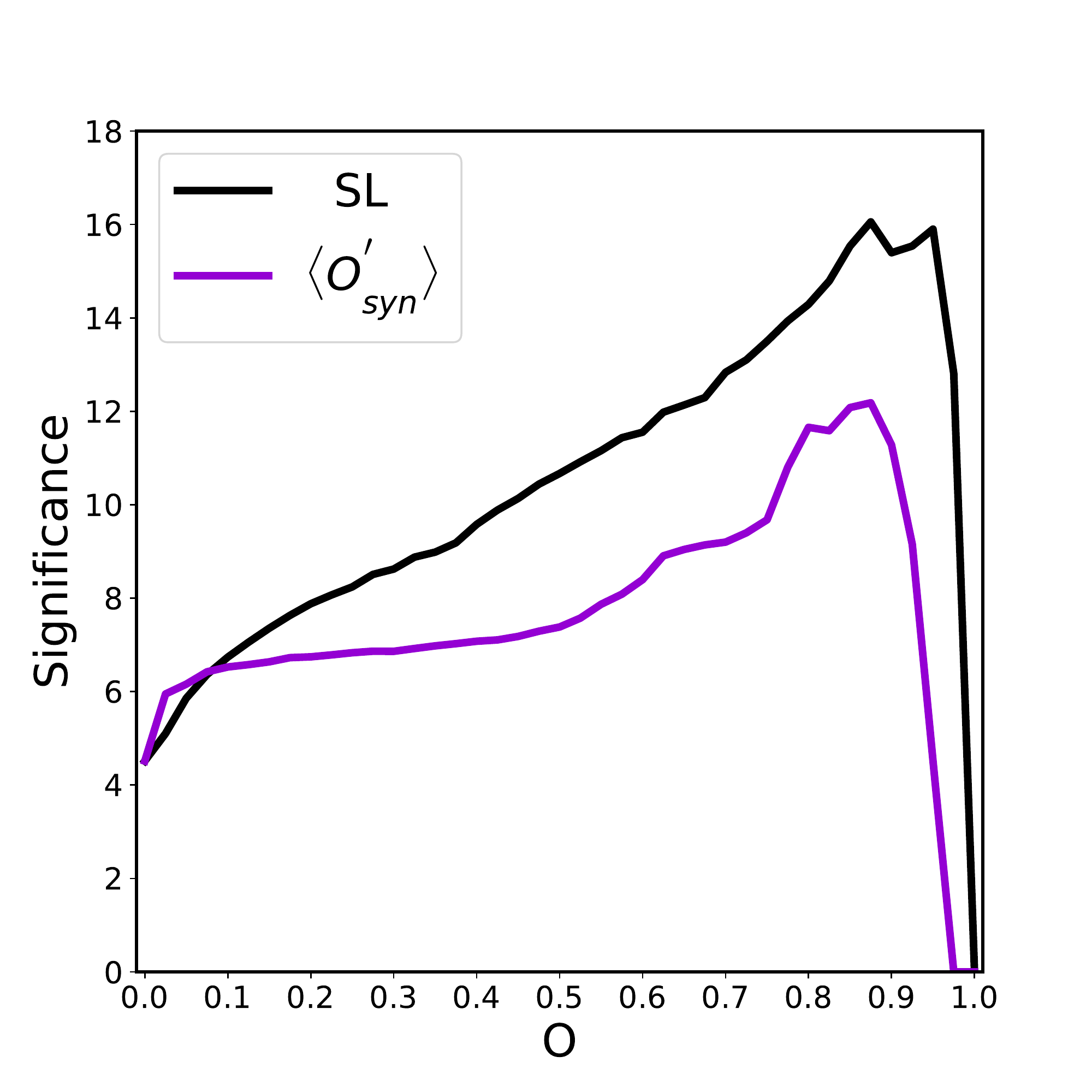}
\caption{$\tilde t \bar{\tilde t}$}
\label{}
\end{subfigure}
\caption{Statistical significance (assuming an integrated luminosity of 3 ab$^{-1}$ at $\sqrt{s}=13$ TeV) as a function of the $\langle \mathcal{O'}_{syn} \rangle$ threshold (40 bins are defined) in the $t\bar t \gamma\gamma$ analysis. The signal events considered correspond to (a) $t\bar t h$ and (b) $\tilde t \bar{\tilde t}$. As a reference, the SL performance has also been reported.}
\label{fig:physsig}
\end{figure}  

The ROC curves and their AUC values for the set of novelty evaluators considered are shown in \fig{physauc}. As previously discussed, $\mathcal{O}'_{\rm iso}$ and $\mathcal{O}_{\rm clu}$ are sensitive to one of the two signal patterns (resonance or continuous spectrum) only, whereas $\mathcal{O'}_{\rm syn}$ brings in a further improvement, performing more stably in both cases. Finally, we show the statistical significances for discovery of both types of signals in \fig{physsig}. The sensitivity from a supervised-learning classifier is also shown for each case. It represents the best performance which could be achieved. Encouragingly, the sensitivities of $\langle \mathcal{O'}_{\rm syn} \rangle$ are not far from the ``ideal'' ones. Besides, these sensitivities are compatible with the results of real-data analyses. The  $t\bar t h$ production with $h\to \gamma\gamma$ was analyzed in Ref.~\cite{Aaboud:2018xdt} using $\unit[36.1]{fb^{-1}}$ of ATLAS data at $\sqrt{s}=13$ TeV. This analysis yielded a significance of 1.16$\sigma$ in the \emph{$t\bar{t}H$}-lep region, with the events in the di-photon mass window containing 90\% signals (see Table 27 in Ref.~\cite{Aaboud:2018xdt}).\footnote{The ATLAS analysis using $\unit[139]{fb^{-1}}$ of data in Ref.~\cite{ATLAS:2019aqa} yielded a significance of 4.9$\sigma$ against the background-only hypothesis. Notably, the \emph{$t\bar{t}H$}-lep region defined in this paper is different from the one in Ref.~\cite{Aaboud:2018xdt}, which employs a looser preselection.} As a comparison, Fig.~\ref{fig:physsig}(a) indicates an optimal significance $\gtrsim 1.0 \sigma$ with the same amount of integrated luminosity. Separately, the stop pair production with a full decay into $t\bar{t}\gamma\gamma + 2\tilde{G}$ was analyzed in Ref.~\cite{Sirunyan:2019hzr}, using $\unit[35.9]{fb^{-1}}$ of CMS data at $\sqrt{s}=13$ TeV. According to Fig.~5 in Ref.~\cite{Sirunyan:2019hzr}, the benchmark scenario studied here ($i.e.$ $m_{\tilde t}=1$ TeV and $m_{\chi}=0.2$ TeV) has been excluded with a significance $> 2.0 \sigma$, considering inclusive decays of $\tilde{\chi}^0_1$ with ${\rm BR}(\tilde{\chi}^0_1\to \gamma \tilde{G}) = {\rm BR}(\tilde{\chi}^0_1\to Z \tilde{G}) =50\%$. Assuming the same luminosity and BR$(t\bar{t}\gamma\gamma + 2\tilde{G})=25\%$, Fig.~\ref{fig:physsig}(b) implies an optimal significance (against the background-only hypothesis instead) of $\sim 3.1 \sigma$. 

\section{Conclusions}
\label{sec:S4}
The null results of the broad program of searches at the LHC so far may imply that the NP signals at colliders could take forms highly unexpected. This strongly motivates to develop strategies that could allow the NP to be detected in a less model-dependent way and with a broader coverage in theory space, hence complementing the current model-dependent search programs at the LHC. The ML techniques of novelty detection can well serve this purpose, since they are essentially designed to detect novel events without a prior knowledge. 

Following Ref.~\cite{Hajer:2018kqm}, where $\mathcal{O}_{\rm syn}$ ($i.e.$ a novelty evaluator utilizing the complementarity between the $k$-NN based $\mathcal{O}_{\rm iso}$ and $\mathcal{O}_{\rm clu}$ evaluators) was proposed to improve the detection sensitivity, we have develop an analysis scheme to make use of this feature in a more systematic way. One improvement here is that we adopt an additional step to resort the bins of ${\mathcal O_{\rm iso}}$ (see Step III in Sec.~\ref{subsec:design}), which defines $\mathcal{O}'_{\rm iso}$, according to the level of deviation of the testing sample from the reference sample in its each bin.  As discussed, the signal events from the same region of the feature space tend to have close ${\mathcal O_{\rm iso}}$ scores. These events however may not be well-identified by ${\mathcal O_{\rm iso}}$, if they are from the background bulk. This step to a large extent resolves this problem. Then we can define $\mathcal{O}_{\rm syn}$ based on the $\mathcal{O}'_{\rm iso}$ and $\mathcal{O}_{\rm clu}$ scores. To interpret the confidence level of the data deviation based on $\mathcal{O}_{\rm syn}$ with Gaussian/Poisson statistics, we select a signal region and background region separated by a certain $\mathcal{O}_{\rm syn}$ threshold. Then we use the latent space as inputs for a supervised DNN to classify these two regions. The output score of this DNN defines our final novelty score $\mathcal{O'}_{\rm syn}$ to quantify the level of deviation of the data from known patterns. Such a treatment improves the confidence level of $\mathcal{O}_{\rm syn}$ in a general context. If the data statistics is sufficiently high, one can even use the full final-state particle kinematics as input to separate the signal and non-signal regions. Then the effect of information loss caused by dimensionality reduction will be diminished. 
Here we would stress that this scheme is rather broad. It represents a class of designs, where the $k$NN based $\mathcal{O}_{\rm iso}$ and $\mathcal{O}_{\rm clu}$ can be replaced with another pair of isolation- and clustering-based evaluators (see Table~\ref{tab:algorithms}), with the features of $\mathcal{O}_{\rm syn}$ and $\mathcal{O}'_{\rm syn}$ being qualitatively unchanged. 

We then conduct a comparative study on novelty evaluators, and demonstrate the generality and efficiency of $\mathcal{O'}_{\rm syn}$ in a variety of NP scenarios which are mimicked with two dimensional Gaussian samples. The yielded signatures range from loose clustering in the center of the known-pattern distribution to compact isolation. We subsequently apply this study to the LHC detection of the SM $t\bar t h$ production and the direct stop-quark pair production in gravity-mediated SUSY as novel events in the $t\bar{t}\gamma\gamma$ channel. These two scenarios yield signal patterns with a sharp resonance and a broad distribution of $m_{\gamma\gamma}$, respectively. With $\mathcal{O'}_{\rm syn}$, we successfully identify both types of signal, reaching a discover/exclusion confidence level comparable to the dedicated supervised-learning searches. We would like to stress: although the sensitivity for these two physical cases might depend on the chosen latent-space dimensionality using the invariant-mass-preserving AE architecture, a systematic way to define the symmetry-preserving latent space for novelty evaluation is another broad aspect that deserves further exploration (see e.g. Ref.~\cite{DBLP:journals/corr/abs-1902-04615} in this direction).

Our study so far is semi-supervised in the sense that the training or reference samples are expected to be generated using the MC simulation tools. Such a method will unavoidably introduce some systematics including QCD uncertainties due to the inaccuracy of simulation. One way to reduce such systematics could be developing adversarially-trained autoencoders where the sensitivity of autoencoders to the simulation-caused bias is suppressed~\cite{Blance:2019ibf}. Alternatively, one can extend semi-supervised learning to fully unsupervised learning or a data-driven method. To achieve this, one needs to extrapolate the backgrounds in the signal region from the control regions. This strategy may reduce the systematics to be below 10\% (for some example using the ABCD method, see, e.g.,~\cite{CMS:2018mts}). The background extrapolation could be further improved by using a generative adversarial network where a background sample will be generated to mimic the data. Notably, in these processes the novelty evaluation for collider events will be essentially unchanged. We leave these explorations to a future work.

\section*{Acknowledgements}
We would like to thank Yanjun Tu for useful discussions. T.L. was supported by the General Research Fund under Grant No. 16304315, which was issued by the Research Grants Council of Hong Kong S.A.R.  Fermilab is operated by Fermi Research Alliance, LLC under  contract number DE-AC02-07CH11359  with  the United States Department of Energy. A.J. was supported in part by the Spanish Ministerio de Econom\'ia y Competitividad under projects RTI2018-096930-B-I00 and Centro de Excelencia Severo Ochoa SEV-2016-0588.

\newpage

\appendix
\section{More Figures}
\label{sec:AA}

\begin{figure}[ht]
\centering\includegraphics[width=1.0 \textwidth]{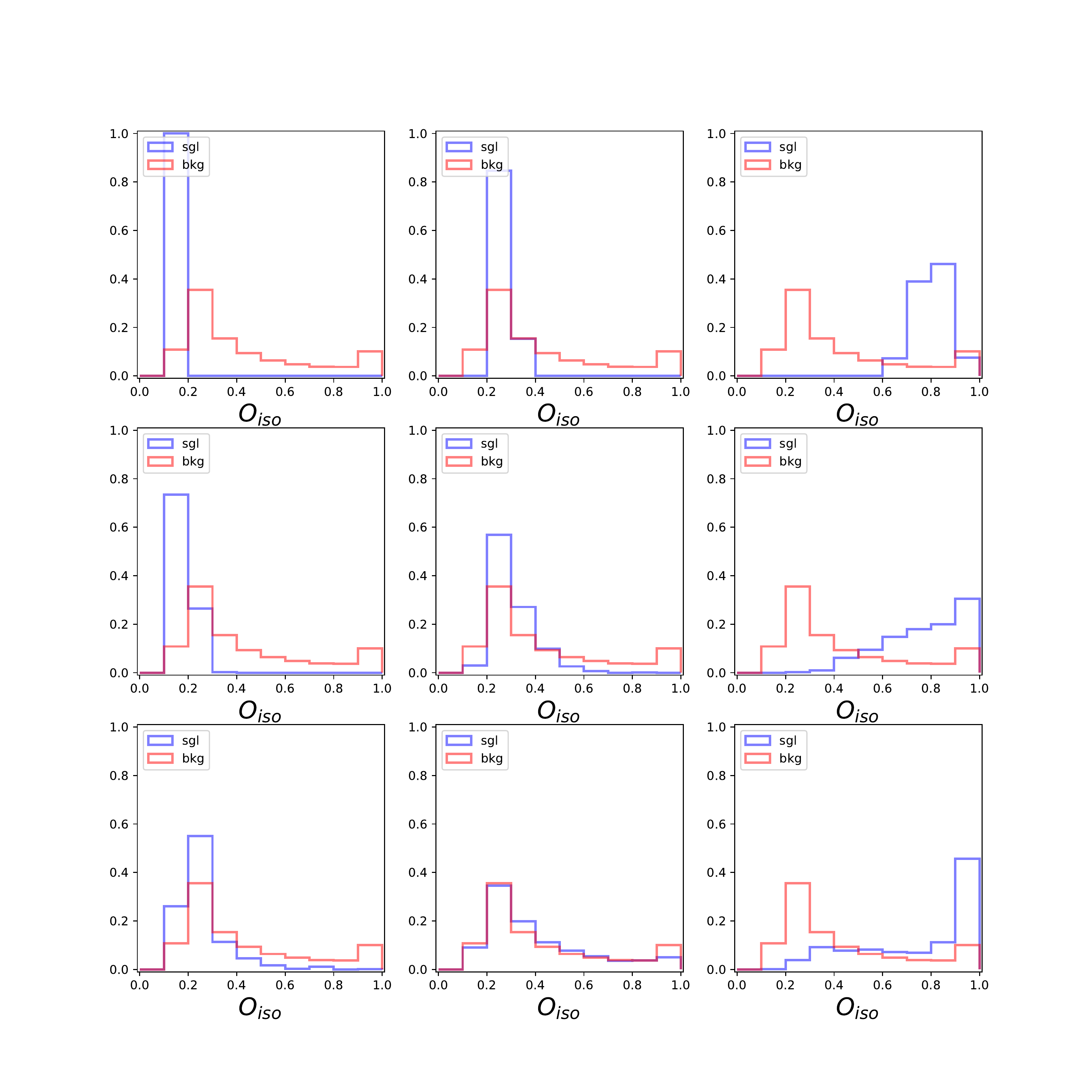}
\caption{Novelty score according to $O_{\rm iso}$ for the 2D Gaussian testing samples corresponding to the same BPs as in~\fig{gaussian}.}
\label{fig:oiso}
\end{figure} 

\begin{figure}[ht]
\centering\includegraphics[width=1.0 \textwidth]{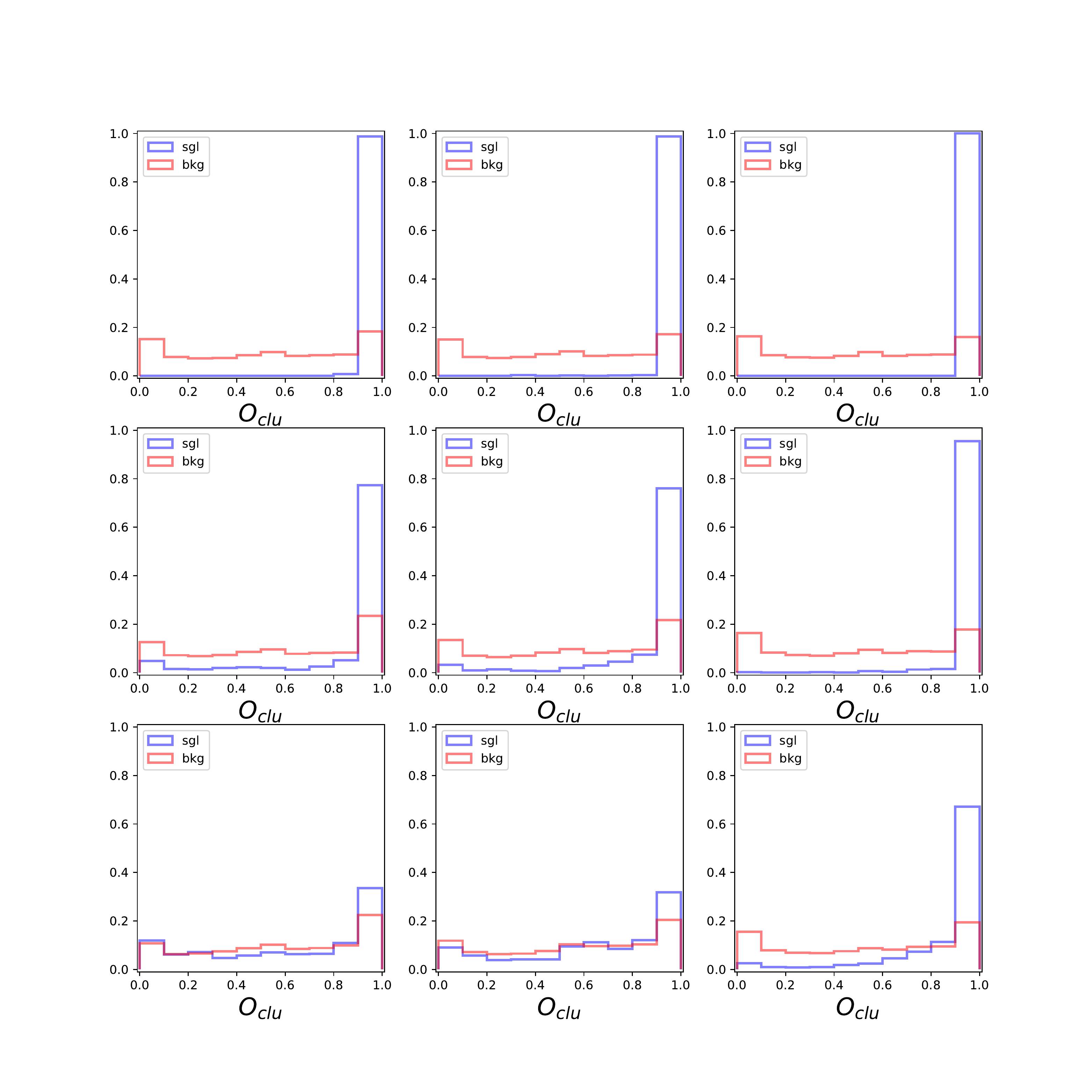}
\caption{Novelty score according to $O_{\rm clu}$ for the 2D Gaussian testing samples corresponding to the same BPs as in~\fig{gaussian}.}
\label{fig:oclu}
\end{figure}

\clearpage
\bibliographystyle{JHEP}
\bibliography{ref}

\providecommand{\href}[2]{#2}\begingroup\raggedright\begin{thebibliography}{10}

\bibitem{Hajer:2018kqm}
J.~Hajer, Y.-Y. Li, T.~Liu, and H.~Wang, {\it {Novelty Detection Meets Collider
  Physics}},  {\em Phys. Rev. D} {\bf 101} (2020) 076015,
  [\href{http://arxiv.org/abs/1807.10261}{{\tt arXiv:1807.10261}}].

\bibitem{DAgnolo:2019vbw}
R.~T. D'Agnolo, G.~Grosso, M.~Pierini, A.~Wulzer, and M.~Zanetti, {\it
  {Learning multivariate new physics}},  {\em Eur. Phys. J. C} {\bf 81} (2021)
  89, [\href{http://arxiv.org/abs/1912.12155}{{\tt arXiv:1912.12155}}].

\bibitem{Guest:2018yhq}
D.~Guest, K.~Cranmer, and D.~Whiteson, {\it {Deep Learning and its Application
  to LHC Physics}},  {\em Ann. Rev. Nucl. Part. Sci.} {\bf 68} (2018) 161,
  [\href{http://arxiv.org/abs/1806.11484}{{\tt arXiv:1806.11484}}].

\bibitem{1996}
P.~C. Bhat, {\it {Search for the Top Quark at D0 using multivariate methods}},
  {\em AIP Conference Proceedings} (1996).

\bibitem{PRLD0}
{D\O{} Collaboration}, {\it Evidence for production of single top quarks and
  first direct measurement of $|{V}_{tb}|$},  {\em Phys. Rev. Lett.} {\bf 98}
  (2007) 181802.

\bibitem{cios2017deep}
K.~J. Cios, {\it {Deep Neural Networks - A Brief History}},
  \href{http://arxiv.org/abs/1701.05549}{{\tt arXiv:1701.05549}}.

\bibitem{boosting}
R.~Schapire, {\it The strength of weak learnability},  {\em Machine Learning}
  {\bf 5} (1990) 197.

\bibitem{Roe:2004na}
B.~P. Roe, H.-J. Yang, J.~Zhu, Y.~Liu, I.~Stancu, and G.~McGregor, {\it
  {Boosted decision trees, an alternative to artificial neural networks}},
  {\em Nucl. Instrum. Meth. A} {\bf 543} (2005) 577,
  [\href{http://arxiv.org/abs/physics/0408124}{{\tt physics/0408124}}].

\bibitem{Aad:2012tfa}
{ATLAS Collaboration}, {\it {Observation of a new particle in the search for
  the Standard Model Higgs boson with the ATLAS detector at the LHC}},  {\em
  Phys. Lett. B} {\bf 716} (2012) 1,
  [\href{http://arxiv.org/abs/1207.7214}{{\tt arXiv:1207.7214}}].

\bibitem{Chatrchyan:2012xdj}
{CMS Collaboration}, {\it {Observation of a New Boson at a Mass of 125 GeV with
  the CMS Experiment at the LHC}},  {\em Phys. Lett. B} {\bf 716} (2012) 30,
  [\href{http://arxiv.org/abs/1207.7235}{{\tt arXiv:1207.7235}}].

\bibitem{Pimentel14}
M.~A. Pimentel, D.~A. Clifton, L.~Clifton, and L.~Tarassenko, {\it A review of
  novelty detection},  {\em Signal Processing} {\bf 99} (2014) 215.

\bibitem{DAgnolo:2018cun}
R.~T. D'Agnolo and A.~Wulzer, {\it {Learning New Physics from a Machine}},
  {\em Phys. Rev. D} {\bf 99} (2019) 015014,
  [\href{http://arxiv.org/abs/1806.02350}{{\tt arXiv:1806.02350}}].

\bibitem{DeSimone:2018efk}
A.~De~Simone and T.~Jacques, {\it {Guiding New Physics Searches with
  Unsupervised Learning}},  {\em Eur. Phys. J. C} {\bf 79} (2019) 289,
  [\href{http://arxiv.org/abs/1807.06038}{{\tt arXiv:1807.06038}}].

\bibitem{Heimel:2018mkt}
T.~Heimel, G.~Kasieczka, T.~Plehn, and J.~M. Thompson, {\it {QCD or What?}},
  {\em SciPost Phys.} {\bf 6} (2019) 030,
  [\href{http://arxiv.org/abs/1808.08979}{{\tt arXiv:1808.08979}}].

\bibitem{Farina:2018fyg}
M.~Farina, Y.~Nakai, and D.~Shih, {\it {Searching for New Physics with Deep
  Autoencoders}},  {\em Phys. Rev. D} {\bf 101} (2020) 075021,
  [\href{http://arxiv.org/abs/1808.08992}{{\tt arXiv:1808.08992}}].

\bibitem{Roy:2019jae}
T.~S. Roy and A.~H. Vijay, {\it {A robust anomaly finder based on
  autoencoders}},  \href{http://arxiv.org/abs/1903.02032}{{\tt
  arXiv:1903.02032}}.

\bibitem{Cerri:2018anq}
O.~Cerri, T.~Q. Nguyen, M.~Pierini, M.~Spiropulu, and J.-R. Vlimant, {\it
  {Variational Autoencoders for New Physics Mining at the Large Hadron
  Collider}},  {\em JHEP} {\bf 05} (2019) 036,
  [\href{http://arxiv.org/abs/1811.10276}{{\tt arXiv:1811.10276}}].

\bibitem{Cheng:2020dal}
T.~Cheng, J.-F. Arguin, J.~Leissner-Martin, J.~Pilette, and T.~Golling, {\it
  {Variational Autoencoders for Anomalous Jet Tagging}},
  \href{http://arxiv.org/abs/2007.01850}{{\tt arXiv:2007.01850}}.

\bibitem{Knapp:2020dde}
O.~Knapp, O.~Cerri, G.~Dissertori, T.~Q. Nguyen, M.~Pierini, and J.-R. Vlimant,
  {\it {Adversarially Learned Anomaly Detection on CMS Open Data:
  re-discovering the top quark}},  {\em Eur. Phys. J. Plus} {\bf 136} (2021)
  236, [\href{http://arxiv.org/abs/2005.01598}{{\tt arXiv:2005.01598}}].

\bibitem{Blance:2019ibf}
A.~Blance, M.~Spannowsky, and P.~Waite, {\it {Adversarially-trained
  autoencoders for robust unsupervised new physics searches}},  {\em JHEP} {\bf
  10} (2019) 047, [\href{http://arxiv.org/abs/1905.10384}{{\tt
  arXiv:1905.10384}}].

\bibitem{Bortolato:2021zic}
B.~Bortolato, B.~M. Dillon, J.~F. Kamenik, and A.~Smolkovi\v{c}, {\it {Bump
  Hunting in Latent Space}},  \href{http://arxiv.org/abs/2103.06595}{{\tt
  arXiv:2103.06595}}.

\bibitem{Atkinson:2021nlt}
O.~Atkinson, A.~Bhardwaj, C.~Englert, V.~S. Ngairangbam, and M.~Spannowsky,
  {\it {Anomaly detection with Convolutional Graph Neural Networks}},  {\em
  JHEP} {\bf 08} (2021) 080, [\href{http://arxiv.org/abs/2105.07988}{{\tt
  arXiv:2105.07988}}].

\bibitem{Ngairangbam:2021yma}
V.~S. Ngairangbam, M.~Spannowsky, and M.~Takeuchi, {\it {Anomaly detection in
  high-energy physics using a quantum autoencoder}},
  \href{http://arxiv.org/abs/2112.04958}{{\tt arXiv:2112.04958}}.

\bibitem{Mullin:2019mmh}
A.~Mullin, S.~Nicholls, H.~Pacey, M.~Parker, M.~White, and S.~Williams, {\it
  {Does SUSY have friends? A new approach for LHC event analysis}},  {\em JHEP}
  {\bf 02} (2021) 160, [\href{http://arxiv.org/abs/1912.10625}{{\tt
  arXiv:1912.10625}}].

\bibitem{Buss:2022lxw}
T.~Buss, B.~M. Dillon, T.~Finke, M.~Kr\"amer, A.~Morandini, A.~M\"uck,
  I.~Oleksiuk, and T.~Plehn, {\it {What's Anomalous in LHC Jets?}},
  \href{http://arxiv.org/abs/2202.00686}{{\tt arXiv:2202.00686}}.

\bibitem{Chakravarti:2021svb}
P.~Chakravarti, M.~Kuusela, J.~Lei, and L.~Wasserman, {\it {Model-Independent
  Detection of New Physics Signals Using Interpretable Semi-Supervised
  Classifier Tests}},  \href{http://arxiv.org/abs/2102.07679}{{\tt
  arXiv:2102.07679}}.

\bibitem{dAgnolo:2021aun}
R.~T. d'Agnolo, G.~Grosso, M.~Pierini, A.~Wulzer, and M.~Zanetti, {\it
  {Learning new physics from an imperfect machine}},  {\em Eur. Phys. J. C}
  {\bf 82} (2022), no.~3 275, [\href{http://arxiv.org/abs/2111.13633}{{\tt
  arXiv:2111.13633}}].

\bibitem{Aguilar-Saavedra:2021utu}
J.~A. Aguilar-Saavedra, {\it {Anomaly detection from mass unspecific jet
  tagging}},  {\em Eur. Phys. J. C} {\bf 82} (2022), no.~2 130,
  [\href{http://arxiv.org/abs/2111.02647}{{\tt arXiv:2111.02647}}].

\bibitem{Nachman:2020lpy}
B.~Nachman and D.~Shih, {\it {Anomaly Detection with Density Estimation}},
  {\em Phys. Rev. D} {\bf 101} (2020) 075042,
  [\href{http://arxiv.org/abs/2001.04990}{{\tt arXiv:2001.04990}}].

\bibitem{Alvarez:2021zje}
E.~Alvarez, M.~Spannowsky, and M.~Szewc, {\it {Unsupervised quark/gluon jet
  tagging with Poissonian Mixture Models}},
  \href{http://arxiv.org/abs/2112.11352}{{\tt arXiv:2112.11352}}.

\bibitem{Metodiev:2017vrx}
E.~M. Metodiev, B.~Nachman, and J.~Thaler, {\it {Classification without labels:
  Learning from mixed samples in high energy physics}},  {\em JHEP} {\bf 10}
  (2017) 174, [\href{http://arxiv.org/abs/1708.02949}{{\tt arXiv:1708.02949}}].

\bibitem{Collins:2018epr}
J.~H. Collins, K.~Howe, and B.~Nachman, {\it {Anomaly Detection for Resonant
  New Physics with Machine Learning}},  {\em Phys. Rev. Lett.} {\bf 121} (2018)
  241803, [\href{http://arxiv.org/abs/1805.02664}{{\tt arXiv:1805.02664}}].

\bibitem{Collins:2019jip}
J.~H. Collins, K.~Howe, and B.~Nachman, {\it {Extending the search for new
  resonances with machine learning}},  {\em Phys. Rev. D} {\bf 99} (2019)
  014038, [\href{http://arxiv.org/abs/1902.02634}{{\tt arXiv:1902.02634}}].

\bibitem{Aad:2020cws}
{ATLAS Collaboration}, {\it {Dijet resonance search with weak supervision using
  $\sqrt{s}=13$ TeV $pp$ collisions in the ATLAS detector}},  {\em Phys. Rev.
  Lett.} {\bf 125} (2020) 131801, [\href{http://arxiv.org/abs/2005.02983}{{\tt
  arXiv:2005.02983}}].

\bibitem{Amram:2020ykb}
O.~Amram and C.~M. Suarez, {\it {Tag N\textquoteright{} Train: a technique to
  train improved classifiers on unlabeled data}},  {\em JHEP} {\bf 01} (2021)
  153, [\href{http://arxiv.org/abs/2002.12376}{{\tt arXiv:2002.12376}}].

\bibitem{Andreassen:2020nkr}
A.~Andreassen, B.~Nachman, and D.~Shih, {\it {Simulation Assisted
  Likelihood-free Anomaly Detection}},  {\em Phys. Rev. D} {\bf 101} (2020)
  095004, [\href{http://arxiv.org/abs/2001.05001}{{\tt arXiv:2001.05001}}].

\bibitem{Dahbi:2020zjw}
S.-e. Dahbi, J.~Choma, B.~Mellado, G.~Mokgatitswane, X.~Ruan, T.~Celik, and
  B.~Lieberman, {\it {Machine learning approach for the search of resonances
  with topological features at the Large Hadron Collider}},
  \href{http://arxiv.org/abs/2011.09863}{{\tt arXiv:2011.09863}}.

\bibitem{Mikuni:2020qds}
V.~Mikuni and F.~Canelli, {\it {Unsupervised clustering for collider physics}},
   {\em Phys. Rev. D} {\bf 103} (2021) 092007,
  [\href{http://arxiv.org/abs/2010.07106}{{\tt arXiv:2010.07106}}].

\bibitem{KL-D}
S.~Kullback and R.~A. Leibler, {\it {On Information and Sufficiency}},  {\em
  The Annals of Mathematical Statistics} {\bf 22} (1951) 79.

\bibitem{Komiske:2019fks}
P.~T. Komiske, E.~M. Metodiev, and J.~Thaler, {\it {Metric Space of Collider
  Events}},  {\em Phys. Rev. Lett.} {\bf 123} (2019) 041801,
  [\href{http://arxiv.org/abs/1902.02346}{{\tt arXiv:1902.02346}}].

\bibitem{NIPS2017_6c1da886}
G.~Papamakarios, T.~Pavlakou, and I.~Murray, {\it Masked autoregressive flow
  for density estimation},  in {\em Advances in Neural Information Processing
  Systems} (I.~Guyon, U.~V. Luxburg, S.~Bengio, H.~Wallach, R.~Fergus,
  S.~Vishwanathan, and R.~Garnett, eds.), vol.~30, Curran Associates, Inc.,
  2017.

\bibitem{10.1093/nsr/nwx106}
Z.-H. Zhou, {\it {A brief introduction to weakly supervised learning}},  {\em
  National Science Review} {\bf 5} (08, 2017) 44,
  [\href{http://arxiv.org/abs/https://academic.oup.com/nsr/article-pdf/5/1/44/31567770/nwx106.pdf}{{\tt
  https://academic.oup.com/nsr/article-pdf/5/1/44/31567770/nwx106.pdf}}].

\bibitem{Vincent08}
P.~Vincent, H.~Larochelle, Y.~Bengio, and P.-A. Manzagol, {\it Extracting and
  composing robust features with denoising autoencoders},  in {\em Proceedings
  of the 25th International Conference on Machine Learning}, ICML '08, (New
  York, NY, USA), pp.~1096--1103, ACM, 2008.

\bibitem{chollet2015keras}
F.~Chollet et~al., ``Keras.'' \url{https://keras.io}, 2015.

\bibitem{Collins:2021nxn}
J.~H. Collins, P.~Mart\'\i{}n-Ramiro, B.~Nachman, and D.~Shih, {\it {Comparing
  weak- and unsupervised methods for resonant anomaly detection}},  {\em Eur.
  Phys. J. C} {\bf 81} (2021) 617, [\href{http://arxiv.org/abs/2104.02092}{{\tt
  arXiv:2104.02092}}].

\bibitem{Caron:2021wmq}
S.~Caron, L.~Hendriks, and R.~Verheyen, {\it {Rare and Different: Anomaly
  Scores from a combination of likelihood and out-of-distribution models to
  detect new physics at the LHC}},  \href{http://arxiv.org/abs/2106.10164}{{\tt
  arXiv:2106.10164}}.

\bibitem{Dillon:2021nxw}
B.~M. Dillon, T.~Plehn, C.~Sauer, and P.~Sorrenson, {\it {Better Latent Spaces
  for Better Autoencoders}},  {\em SciPost Phys.} {\bf 11} (2021) 061,
  [\href{http://arxiv.org/abs/2104.08291}{{\tt arXiv:2104.08291}}].

\bibitem{Finke:2021sdf}
T.~Finke, M.~Kr\"amer, A.~Morandini, A.~M\"uck, and I.~Oleksiyuk, {\it
  {Autoencoders for unsupervised anomaly detection in high energy physics}},
  {\em JHEP} {\bf 06} (2021) 161, [\href{http://arxiv.org/abs/2104.09051}{{\tt
  arXiv:2104.09051}}].

\bibitem{Alwall:2014hca}
J.~Alwall, R.~Frederix, S.~Frixione, V.~Hirschi, F.~Maltoni, O.~Mattelaer,
  H.~S. Shao, T.~Stelzer, P.~Torrielli, and M.~Zaro, {\it {The automated
  computation of tree-level and next-to-leading order differential cross
  sections, and their matching to parton shower simulations}},  {\em JHEP} {\bf
  07} (2014) 079, [\href{http://arxiv.org/abs/1405.0301}{{\tt
  arXiv:1405.0301}}].

\bibitem{Sjostrand:2007gs}
T.~Sjostrand, S.~Mrenna, and P.~Z. Skands, {\it {A Brief Introduction to PYTHIA
  8.1}},  {\em Comput. Phys. Commun.} {\bf 178} (2008) 852,
  [\href{http://arxiv.org/abs/0710.3820}{{\tt arXiv:0710.3820}}].

\bibitem{deFavereau:2013fsa}
J.~de~Favereau, C.~Delaere, P.~Demin, A.~Giammanco, V.~Lemaître, A.~Mertens,
  and M.~Selvaggi, {\it {DELPHES 3, A modular framework for fast simulation of
  a generic collider experiment}},  {\em JHEP} {\bf 02} (2014) 057,
  [\href{http://arxiv.org/abs/1307.6346}{{\tt arXiv:1307.6346}}].

\bibitem{ATLAS:2019aqa}
{ATLAS Collaboration}, {\it {Measurement of Higgs boson production in
  association with a $t\overline t$ pair in the diphoton decay channel using
  139~fb$^{-1}$ of LHC data collected at $\sqrt{s} = 13$~TeV by the ATLAS
  experiment}},  {\em ATLAS-CONF-2019-004} (2019).
  \url{http://cdsweb.cern.ch/record/2668103}.

\bibitem{Aaboud:2018xdt}
{ATLAS Collaboration}, {\it {Measurements of Higgs boson properties in the
  diphoton decay channel with 36 fb$^{-1}$ of $pp$ collision data at $\sqrt{s}
  = 13$ TeV with the ATLAS detector}},  {\em Phys. Rev. D} {\bf 98} (2018)
  052005, [\href{http://arxiv.org/abs/1802.04146}{{\tt arXiv:1802.04146}}].

\bibitem{Homiller:2018dgu}
S.~Homiller and P.~Meade, {\it {Measurement of the Triple Higgs Coupling at a
  HE-LHC}},  {\em JHEP} {\bf 03} (2019) 055,
  [\href{http://arxiv.org/abs/1811.02572}{{\tt arXiv:1811.02572}}].

\bibitem{Borschensky:2014cia}
C.~Borschensky, M.~Krämer, A.~Kulesza, M.~Mangano, S.~Padhi, T.~Plehn, and
  X.~Portell, {\it {Squark and gluino production cross sections in pp
  collisions at $\sqrt{s}$ = 13, 14, 33 and 100 TeV}},  {\em Eur. Phys. J. C}
  {\bf 74} (2014) 3174, [\href{http://arxiv.org/abs/1407.5066}{{\tt
  arXiv:1407.5066}}].

\bibitem{zeiler2012adadelta}
M.~D. Zeiler, {\it Adadelta: An adaptive learning rate method},
  \href{http://arxiv.org/abs/1212.5701}{{\tt arXiv:1212.5701}}.

\bibitem{Meng_2017}
Q.~Meng, D.~Catchpoole, D.~Skillicom, and P.~J. Kennedy, {\it Relational
  autoencoder for feature extraction},  {\em 2017 International Joint
  Conference on Neural Networks (IJCNN)} (2017) 364.

\bibitem{Sirunyan:2019hzr}
{CMS Collaboration}, {\it {Search for supersymmetry in events with a photon,
  jets, $\mathrm {b}$ -jets, and missing transverse momentum in proton–proton
  collisions at 13 $\,\text {Te}\text {V}$}},  {\em Eur. Phys. J. C} {\bf 79}
  (2019) 444, [\href{http://arxiv.org/abs/1901.06726}{{\tt arXiv:1901.06726}}].

\bibitem{DBLP:journals/corr/abs-1902-04615}
T.~S. Cohen, M.~Weiler, B.~Kicanaoglu, and M.~Welling, {\it Gauge equivariant
  convolutional networks and the icosahedral {CNN}},  {\em CoRR} {\bf
  abs/1902.04615} (2019) [\href{http://arxiv.org/abs/1902.04615}{{\tt
  arXiv:1902.04615}}].

\bibitem{CMS:2018mts}
{\bf CMS} Collaboration, A.~M. Sirunyan et~al., {\it {Search for pair-produced
  resonances decaying to quark pairs in proton-proton collisions at $\sqrt{s}=$
  13 TeV}},  {\em Phys. Rev. D} {\bf 98} (2018), no.~11 112014,
  [\href{http://arxiv.org/abs/1808.03124}{{\tt arXiv:1808.03124}}].

\end{thebibliography}\endgroup

\end{document}